\documentclass{article}
\input epsf.sty
\input colordvi.sty

\newcommand{\half}{\frac{1}{2}}
\newcommand{\bright}{\begin{flushright}}
\newcommand{\eright}{\end{flushright}}
\newcommand{\bminip}{\begin{minipage}}
\newcommand{\eminip}{\end{minipage}}
\newcommand{\bcent}{\begin{center}}
\newcommand{\ecent}{\end{center}}
\newcommand{\beq}{\begin{equation}}
\newcommand{\eeq}{\end{equation}}
\newcommand{\beqa}{\begin{eqnarray}}
\newcommand{\eeqa}{\end{eqnarray}}
\newcommand{\barr}{\begin{array}}
\newcommand{\earr}{\end{array}}
\renewcommand{\theequation}{\arabic{section}.\arabic{equation}}

\newcommand{\nnb}{\nonumber}
\newcommand{\reflef}{(\ref}
\newcommand{\MP}{M_{\rm P}}
\newcommand{\BBbox}{\mbox{\Large $\sqcap$}\hspace{-1.0em}\mbox{\Large $\sqcup$}}\newcommand{\lmd}{\lambda}
\newcommand{\Lmd}{\Lambda}

\newcommand{\lsim}{\mbox{\raisebox{-.3em}{$\;\stackrel{<}{\sim}\;$}}}

\newcommand{\psibar}{\overline{\psi}}

\begin{document}
\baselineskip=0.6cm

\bcent
{\Large\bf Conformal transformation in the scalar-tensor theory applied to the accelerating universe
}\\
Yasunori Fujii \\
Advanced Research Institute for Science and Engineering, \\[.0em]
Waseda University, 169-8555 Tokyo, Japan\\[.0em]
\ecent
\mbox{}\\[-1.6em]

\baselineskip=0.4cm
\noindent
\bcent
\bminip{14cm}
{\large\bf Abstract}\\[-.2em]

{\small 
The scalar-tensor theory is plagued by nagging questions if different conformal frames, in particular the Jordan and Einstein conformal frames, are equivalent to each other.  As a closely related question, there are opposing views on which of the two conformal frames is physically acceptable.  Reinforcing our previous claims, we offer replies based on a cosmological model of the scalar-tensor theory, believed to be a promising theory for understanding the accelerating universe, as well as today's version of the cosmological constant problem.  Exploiting the advantage that this model admits analytical asymptotic solutions, our argument does not depend on whether the underlying theory is invariant under conformal transformations.  Our argument provides partial support for the claimed ``equivalence," but we also present examples that require more careful analyses exploiting field equations.  We also point out that the Jordan conformal frame is suitable for an interpretation in terms of unification theories in physics, for example, string theory and the Kaluza-Klein approach, while the Einstein conformal frame may be acceptable as a physical conformal frame under two conditions:  (i) the simplest constant $\Lmd$ term in the Lagrangian in the Jordan conformal frame; (ii) the revised form of the conventional Brans-Dicke model based on the validity of weak equivalence principle. 
}
\eminip
\ecent
\mbox{}\\[-.4em]

\baselineskip=0.6cm
\section{Introduction}

The scalar-tensor theory \cite{jordan,bd} appears to provide a promising
approach for understanding the accelerating universe \cite{riessperlm}
and the origin of dark energy, though it requires that we add a
potential to trap the scalar field, which is realized by introducing
another scalar field \cite{cup}.  This theory provides answers to the following two fundamental questions concerning the modern version of the cosmological constant problem. (I) Why is the cosmological constant smaller than the theoretically natural estimate by as much as 120 orders of magnitude?  This is the fine-tuning problem. (II) With a time-independent cosmological constant assumed, why are we so lucky that we are able to observe a one-time event in the entire history of the universe?  This is the coincidence problem.

The fact that we are able to ask these questions itself implies the existence of a theory in which there is a constant $\Lmd$ that is much larger than the observed small value indicated by the accelerating universe.  In order to implement this theoretical idea, we exploit the existence of two conformal frames (CFs), the Jordan CF (JCF) and  the Einstein CF (ECF).  We argue that the JCF provides a representation of the Lagrangian which contains a large $\Lmd$, and this prepares a theoretical path to a fundamental theory, while the field equations derived from another representation of the same Lagrangian expressed in the ECF are directly related to the observation of a small $\Lmd$.  One of the purposes of this article is to supplement and correct or restate some of the accounts published previously \cite{cup} with remarks added to obtain a better understanding of the subject.

In view of the general interest of the problems considered here, we attempt to treat them from as practical a point of view as possible, without considering any details with regard to the question whether or not there exists a conformal {\em invariance} of the underlying theory. In particular, we derive conclusions mainly from the explicit examples derived from the cosmological model, hereafter to be called ST$\Lmd$ cosmology, for the scalar-tensor-$\Lmd$ cosmology, based on the scalar-tensor theory supplemented with a large, constant $\Lambda$.  This simple model has been shown to provide a crucial step toward more realistic approaches to understand details of the accelerating universe \cite{cup}.  It is unlikely that further developments of this approach, for example, by introducing an additional scalar field, will affect the argument given below on the conformal transformation (CFTR), sometimes called Weyl rescaling.

The possible physical significance of the CFTR is discussed in \cite{bd}, and in \cite{dick2}, in which Dicke cited Fierz \cite{fierz}, who quoted Section 28 of \cite{jordan}, in which Jordan described the method for using a mathematical formulation of the CFTR due to Pauli.  Since then, there have been a number of discussions on the nature of the CFTR and CFs, which play unique and indispensable roles in the scalar-tensor theory.  A variety of arguments appear in \cite{falao,falao3}.

Obviously, an underlying question is whether results derived using different CFs are ``equivalent."  It would appear that they would not be equivalent, because we are dealing with theories which are mostly not invariant under CFTR.  However, to definitively answer this question, we need to start by defining what we mean by an equivalence.  We then find certain examples of equivalence, or in plainer language, we find that some results derived from the JCF and the ECF are identical.  But there also exist non-trivial counterexamples.   To elucidate the situation we must go back to the field equations obtained from the Lagrangian expressed in each CF beyond simple dimensional arguments.

In order to offer a quick look at this conclusion before considering details, it seems appropriate here to present a few basic equations and concepts.  We start with the Lagrangian in the JCF, given by \cite{jordan,bd,cup}
\begin{equation}
{\cal L}_{{\rm ST\Lmd}}=\sqrt{-g}\left( \half {\cal M}^2 R -
	\half\epsilon g^{\mu\nu}\partial_{\mu}\phi\partial_{\nu}\phi -
	\Lambda +L_{\rm matter} \right),
\label{cftrI_1}
\end{equation}
where $\phi$ is the scalar field.  The first term on the RHS is the ``nonminimal coupling term" with the spacetime dependent ``gravity mass."  This should be distinguished from the gravitational (as opposed to inertial) mass which is  used more commonly, given by the effective gravitational constant $G_{\rm eff}$ and $\phi$:
\begin{equation}
{\cal M} \left( =(8\pi G_{\rm eff})^{-1/2} \right)=\xi^{1/2}\phi.
\label{cftrI_1a}
\end{equation}
We use the reduced Planck system of units, with $c=\hbar=\MP (=(8\pi G_*)^{-1/2})=1$, where $G_*$ is Newton's constant in ECF, which is assumed to be  an absolute constant, by definition.  In practice, we may replace it by today's value of the observed ``constant" $G$, for which there is no definite evidence of time dependence.  We also assume $c$ and $\hbar$ to be constant throughout this study.  In place of the original symbol $\omega$, we often use $\xi =\epsilon/(4\omega)=1/(4|\omega|)$, with $\epsilon =\pm 1={\rm Sgn}(\omega)$.  According to our sign convention, the kinetic energy of $\phi$ is positive if $\epsilon =+1$.  We choose $\Lmd$ to be a positive, nonzero constant of order 1 in the present system of units.

A CFTR is defined by
\begin{equation}
g_{\mu\nu} \rightarrow g_{*\mu\nu}=\Omega^2(x) g_{\mu\nu},
\label{lc10-2}
\end{equation}
or, equivalently, by
\begin{equation}
ds^2 \rightarrow ds^2_*=\Omega^2(x) ds^2,
\label{lc10-2aa}
\end{equation}
in terms of an arbitrary local function $\Omega (x)$.  The {\em same} Lagrangian, \reflef{cftrI_1}), given in terms of a functional of $g_{\mu\nu}$ can be re-expressed as another functional of $g_{*\mu\nu}$.  The special choice
\begin{equation}
\Omega ={\cal M},
\label{lc10-2a}
\end{equation}
is interesting, because the re-expressed Lagrangian then is given by
\begin{equation}
{\cal L}_{\rm ST\Lmd}\hspace{-.3em}=\hspace{-.3em}\sqrt{-g_{*}}\left(\half R_{*} - {\rm Sgn}(\zeta^2)\half g^{\mu\nu}_{*}\partial_{\mu}\sigma\partial_{\nu}\sigma - V(\sigma)  +L_{\rm *{\rm matter}}  \right),
\label{cftrI_3}
\end{equation}
where  $\sigma$ is a canonical scalar field with $V(\sigma)$ its potential, given by
\begin{eqnarray}
\phi &=&\xi^{-1/2}e^{\zeta \sigma},
\label{cftrI_4} \\
V(\sigma)&=& \Lmd e^{-4\zeta\sigma}. 
\label{cftrI_5}
\end{eqnarray}
Here, $\zeta$ is defined by the well-known equation
\begin{equation}
\zeta^{-2}=6+\epsilon\xi^{-1}= 6+4\omega, 
\label{cftrI_6}
\end{equation}
always assumed to be positive, even for $\epsilon =-1$.  The fact that ${\rm Sgn}(\zeta^2)$ appears in front of the kinetic energy term in \reflef{cftrI_3}) implies that this condition must be imposed in order for the field $\sigma$ to be a non-ghost.

The remarkable point regarding \reflef{cftrI_3})  is that the first term on RHS is the standard Einstein-Hilbert term, due to the relation
\begin{equation}
\Omega^{-2}{\cal M}^2= {\cal M}_*^2 =1,
\label{cftrI_7}
\end{equation}
which defines ${\cal M}_*$ in the ECF.  For this reason, we say that with \reflef{lc10-2}), we move from the JCF to the ECF.  We also say that \reflef{cftrI_1}) and \reflef{cftrI_3}) are two different representations, to be called two different ``Lagrangescapes," of the same theory.

Starting with \reflef{cftrI_1}) or \reflef{cftrI_3}), we develop the ST$\Lmd$ cosmology by specifically assuming a spatially flat Friedmann-Robertson-Walker (FRW) metric, naturally, as well as a spatially uniform scalar field.  We have the scale factors $a$ and $a_*$ in the JCF and the ECF, respectively. They are related by
\begin{equation}
a_* =\Omega a,
\label{lc10-1}
\end{equation}
which is derived easily by applying \reflef{lc10-2}) to $g_{ij}=\delta_{ij}a^2$ and $g_{*ij}=\delta_{ij}a_*^2$ in the spatially flat FRW universe.

Although this model lacks a smooth limit for $\Lmd\rightarrow 0$, it provides an implementation of the scenario of a decaying $\Lmd$, a key to understanding the two questions posed at the beginning of this paper.  Also, remarkably, the model does not suffer from an exponential growth of the universe in spite of the presence of $\Lmd$.   The model also features the advantage that we have attractor solutions that can be obtained analytically.  For this reason, it provides many unambiguous results.

Equations \reflef{cftrI_7}) and \reflef{lc10-1}) show that the scale factor  and the ``gravity length" (i.e. the inverse of the gravity mass) have the same ``conformal dimension," which is equal to 1.  Obviously they are consistent with the length dimension, in a  conventional sense.

Eliminating $\Omega$ from these two equations results in an equation for the corresponding dimensionless combinations,
\begin{equation}
a_* {\cal M}_*=a{\cal M},
\label{lc10_3a}
\end{equation}
which implies that the scale factors of the universe evaluated {\em in
reference to} the gravity lengths are the same in the two CFs, in spite
of the fact that $a$ and $a_*$ themselves behave very differently,
leading occasionally to a somewhat ambiguous example of {\em different physics}.  For this reason, an equation like \reflef{lc10_3a}) might be called a ``conformally dimensionless equation" (CDlessE), which will serve to demonstrate the {\em equivalence} of two CFs.  It is this CDlessE, as we emphasize, that uniquely {\em defines} the concepts of ``equivalence" or ``inequivalence," which have often been used much less unambiguously.  On the other hand, here we may have narrowed the meaning of equivalence. However, we hope that this narrower meaning is convenient at certain levels of application.

In many simple examples, such as those derived in \cite{dick2}, focusing upon dimensionless quantities helps us to correctly analyze physically observable results.  This may not always be the case, however, if we face certain complications that can be dealt with only by going back to the field equations.  We further note that the field equations are derived from a Lagrangian, which can be written down only by specifying the CF.  This may imply the interpretation that fixing a CF determines a specific Lagrangescape.

We also add that choosing a Lagrangian fixes not only dimensionless relations but also much more, for example the canonical scalar field and the potential.  This provides us with many intuitive as well as valuable interpretations of  physical phenomena.  Ultimately a ``theory" is formulated in terms of a Lagrangian.

In addition to the different Lagrangescapes discussed above, there is
another way to characterize a CF, as seen from the statements in
\cite{bd} and \cite{dick2} that CFTR constitutes a local change
of units.  Suppose we are using an atomic clock, for example.  As the
unit of time, we use the frequency of radiation of certain atomic
transitions, which is determined basically by the electron mass.
Strictly speaking, this has to be the reduced electron mass, including a
small contribution from the mass of the nucleus.  For simplicity,
however, we often use the term ``electron mass," without mentioning the
obvious details.  Then, if we use an atomic clock, we have no way of
detecting a change in the mass of the electron.  Indeed, generally
speaking,  we have no way to detect any change in our own time unit.
This obvious fact will be referred to as the ``own-unit-insensitivity principle," according to which using atomic clocks implies that we live in a CF where the electron mass is constant.  In other words, the own-unit-insensitivity principle defines a CF.  We also point out that the same constancy of the electron mass plays a crucial role in determining redshifts of distant objects by comparing the observed atomic spectra with those on the ground.

We do not expect to be capable of discussing every aspect of the issues which can be quite diverse.  We, nevertheless, hope that our careful analyses of rather limited areas will help to broaden our understanding of the subject to more general circumstances.

In sections 2 and 3 we discuss the ST$\Lmd$ cosmology.  Before considering details, however, we make a brief comment on the Brans-Dicke (BD) model \cite{bd}, which has been discussed widely.  Brans and Dicke required the matter Lagrangian $L_{\rm matter}$ given in (1.1) to be decoupled from the scalar field $\phi$ in the JCF, as a way of preserving the weak equivalence principle (WEP) for the following reason.

With the absence of a direct $\phi$-matter coupling in the Lagrangian, the only interaction involving $\phi$ is in the nonminimal coupling term, apart from the minimal coupling taking place through the metric.  We, in fact, find that the nonminimal coupling term includes a mixing interaction between the spacetime-fluctuating part of $\phi$ and the spinless portion of the metric tensor \cite{cup,sant}.  The scalar field does participate in the matter coupling, but only through the metric field.  This implies that the coupling occurs through $T_{\mu\nu}$, the matter energy-momentum tensor, which  embodies WEP in the static limit.  We explicitly obtain the field equation
\begin{equation}
\BBbox\varphi =\zeta^2 \left(T -4\Lmd \right), \quad\mbox{with}\quad \varphi =\half\xi \phi^2 =\half {\cal M}.
\label{BD_1}
\end{equation}
If $L_{\rm matter}$ contains $\phi$, on the other hand, it will possess a direct matter coupling not necessarily confined to $T$, the trace of $T_{\mu\nu}$, as in \reflef{BD_1}).  The matter source of the scalar field may depend on specific properties of the individual matter fields, thus violating WEP.

Also, $T$ vanishes typically for massless matter fields.  In spite of a complication it introduces for scalar fields, this mass dependence is accepted in many practical applications of cosmology, in which the matter distribution is described by a perfect fluid.  In the radiation-dominated universe, in which nonzero particle masses are ignored, we assume that the first term on the RHS of \reflef{BD_1}) vanishes, at least according to the BD model.  In this context, we discuss cosmology in the radiation-dominated universe in Section 2, disregarding details of the BD model.  We propose a revised version of the BD model when we discuss dust-dominated epochs in Section 3.  Even in this revised model, we find it natural to employ the result in Section 2 to be used in more general ST$\Lmd$.

There is another important consequence of the BD model.  To understand this, first note that most of the matter fields come with the mass terms in $L_{\rm matter}$.  The absence of $\phi$ hence implies no time dependence of the masses in JCF:
\begin{equation}
m={\rm const}.
\label{lc10_3b}
\end{equation}
This time independence has a serious consequence in the analysis in Section 3, and it eventually results in a proposed revision of the BD model with $\Lmd$.

Yet another important technical comment regards the manner in which the ST$\Lmd$ cosmology is related to a more realistic approach to the accelerating universe \cite{cup}.   The assumed scalar-field dynamics causes {\em mini-inflations}, which take place {\em sporadically} and are superimposed on the background provided by the ST$\Lmd$ cosmology.  Around the present time, this occurs in the dust-dominated ST$\Lmd$ universe with the scale factor expanding as (time)$^{2/3}$.  This behavior is believed to both precede and follow partially exponential growth, corresponding to the observed acceleration.  In the same way, another mini-inflation must have occurred much earlier, in the radiation-dominated universe, as illustrated explicitly in Fig. 5.8 of \cite{cup}, for example.  In both of these examples, we find the ST$\Lmd$ cosmology to exhibit typical behavior in the radiation- or dust-dominated universe in the conventional sense.

In Section 2, we present two examples which may require some care in using CDlessE, because of the deviation from the simple dimensional argument. Section 3 includes discussion of how the masses of microscopic matter fields are transformed under CFTR, independently of whether these fields respect conformal invariance.   This leads to the required departure from the BD model.  The main conclusions, presented in Section 4 are (a) that the own-unit-insensitivity principle favors the ECF as a physical CF throughout the entire history of the universe and (b)  that the results obtained in an attempt to find the origin of the scalar field favor the specific Lagrangescape realized in the JCF. We also add discussion of the geodesic equations and WEP violation.  Section 5 is devoted to a summary and concluding remarks.

Appendix A provides technical details concerning the relations between the two CFs studied in Subsection 2.2.  In Appendix B, we show that the cosmological term in the initial JCF Lagrangian, \reflef{cftrI_1}), {\em without} modification by the scalar field, is the most likely choice resulting in identifying the ECF as the physical CF, while Appendix C discusses a possible mass term of either $\phi$ or $\sigma$.    In Appendix D, we elaborate a method by which the transformation property of the masses of spinless matter fields under CFTR can be extended to bosonic fields of arbitrary spins, including gauge fields and spinor fields of spin 1/2 and 3/2.  Appendix E gives some details involved in obtaining the JCF solution in the dust-dominated universe, according to the scale-invariant model proposed to replace the BD model.

For convenience, below we provide a list of abbreviations used in this paper.\\[.4em]

\noindent
BD = Brans-Dicke, CDlessE = conformally dimensionless equation, CF = conformal frame, CFTR = conformal transformation, ECF = Einstein conformal frame, FRW = Friedmann-Robertson-Walker, JCF = Jordan conformal frame, LHS = left-hand side, RHS = right-hand side, ST$\Lmd$ cosmology = scalar-tensor-$\Lmd$ cosmology, UFF = universal free-fall, WEP = Weak Equivalence Principle

\section{Cosmology in the radiation-dominated universe}
\setcounter{equation}{0}

In this section, we develop cosmology for the radiation-dominated universe.  In  Subsection 2.1 for the JCF, we emphasize the importance of the solution with $H\!\equiv \!\dot{a}/a\!=\!0$, which is by no means trivial.  We then move on to Subsection 2.2, in which the scale factor happens to behave like $a_*(t_*)=t_*^{1/2}$, as in ordinary analysis.  This suggests that the ECF  be identified with the physical CF.  We also point out that this behavior is rooted in the $\Lmd$ term in \reflef{cftrI_1}), a pure constant that is not modified by $\phi$ as a possible generalization.  In Subsection 2.3, we then present some examples in which the ``equivalence" defined by CDlessE, a relation of the type \reflef{lc10_3a}) or its extension, is violated.

\subsection{Jordan conformal frame}

Consider spatially flat FRW spacetime with a spatially uniform scalar field first in the JCF \cite{dolg}.  The cosmological equations in the radiation-dominated universe are given by (see Chapter 4.4.1 of \cite{cup})
\begin{eqnarray}
6\varphi H^2&=& \half\epsilon \dot{\phi}^2 +\Lmd +\rho -6H\dot{\varphi}, \quad\mbox{where}\quad\varphi =\half \xi \phi^2,\quad\mbox{and} \quad H=\frac{\dot{a}}{a},
\label{dot_1}\\
\ddot{\varphi}+3H\dot{\varphi}&=&4\zeta^2\Lmd,
\label{dot_2}\\
\dot{\rho}+4H\rho &=&0, \label{dot_3}
\end{eqnarray}
where we have dropped the $T$ term on the RHS of \reflef{BD_1}),  on the basis of the BD model, for the moment.  We find the following asymptotic and attractor solution:
\mbox{}\\[-1.9em]
\beqa
a&=& {\rm const}, \label{lc_5} \\
\phi&=& \kappa t,\quad\mbox{with}\quad \kappa = \sqrt{ \frac{4\Lmd}{6\xi+\epsilon}} , 
\label{lc_6}\\
\rho&=& -3\Lmd \frac{2\xi +\epsilon}{6\xi +\epsilon}={\rm const}. \label{lc_7}
\eeqa
In addition to this solution, with $H=0$, there might be other solutions, as well, particularly with $a=t^{1/2}$ for radiation dominance, as suggested by the technical comment made toward the end of preceding section.  The second term on the LHS of \reflef{dot_2}) is now nonzero, and is given by $(3/2)t^{-1}\dot{\varphi}$.  Substituting the same type of solution as \reflef{lc_6}) but with the yet to be determined coefficient $\kappa'$ into \reflef{dot_2}), we obtain
\begin{equation}
\frac{\Lmd}{\kappa^{'2}}=\frac{5}{8}\xi \zeta^{-2}=\frac{5}{8}\xi \left( 6+\epsilon \xi^{-1} \right)= \frac{15}{4}\xi +\frac{5}{8}\epsilon,
\label{lc_6_1}
\end{equation}
where we have used \reflef{cftrI_6}).

From \reflef{dot_3}) we find $\rho \sim t^{-2}$, from which we conclude that we can drop $\rho$ asymptotically on the RHS of \reflef{dot_1}).  Next, we use the same form, $\phi=\kappa' t$, in \reflef{dot_1}), obtaining
\begin{equation}
\frac{\Lmd}{\kappa^{'2}}=\frac{15}{4}\xi -\half \epsilon.
\label{lc_6_2}
\end{equation}
Equating this with \reflef{lc_6_1}), we find no solution for finite $\xi$.  We may instead take the limit $\xi \rightarrow \infty$ or $\omega\rightarrow 0$, which yields $\zeta^2 =1/6$, as discussed by O'Hanlon \cite{ohanlon} in connection with the first proposal of non-Newtonian gravity \cite{yfnonN}, or with a more recent attempt \cite{fiziev}.  Without entering into further details of this exceptional solution, we continue to focus upon the solution with $H=0$, at least for radiation dominance.  However, because a static universe is inconsistent with naive observations, we find it unavoidable to move, probably, to the ECF.

The physical condition $\rho>0$ for \reflef{lc_7}) is satisfied only for 
\begin{equation}
\epsilon =-1,\quad\mbox{and}\quad\frac{1}{6}<\xi < \half,\quad \mbox{and hence }\quad\zeta^2>\frac{1}{4}.
\label{lc_15}
\end{equation}
The first condition suggests the occurrence of a  cancellation between the kinetic energy of $\phi$ and $\Lmd$ on the RHS of the first equation in \reflef{dot_1}), which would suppress the exponential growth of $a$ by relaxing the role of $\Lmd$.

Note that $\epsilon =-1$ implies a ghost nature of $\phi$, but it may not be ruled out immediately.  This is because, as mentioned above, the mixing of the spacetime fluctuating part of $\phi$ with the spinless portion of $g_{\mu\nu}$ caused by the nonminimal coupling term \cite{cup,sant} is removed by introducing the diagonalized field in the weak-field limit at the classical level, with the energy guaranteed to be positive if the condition $\zeta^2 >0$ with \reflef{cftrI_6}) is satisfied, even with $\epsilon =-1$.  A crucial point here is that the presence of  the mixing interaction does not allow us to excite a scalar-field mode without exciting the background spacetime itself from being flat, as explained in Chapters 2.6 and 3.2 (page 70) of \cite{cup}.   From this point of view, it does not seem reasonable to imagine a negative energy in {\em flat} spacetime, thus rejecting JCF with $\epsilon =-1$ categorically as being unphysical \cite{cho}.

The CFTR to the ECF as defined by \reflef{cftrI_1a}) through \reflef{lc10-2a}) is, according to \reflef{lc_6}), given by 
\begin{equation}
\Omega =\xi^{1/2}\phi \left( ={\cal M}\right)=\xi^{1/2}\kappa t.
\label{lc_7b}
\end{equation}

\subsection{Einstein conformal frame}

The field equations in the ECF are (see Chapter 4.4.2 of \cite{cup}) 
\begin{eqnarray}
3H_*^2=\rho_\sigma+\rho_*, &&\hspace{-1em}\mbox{where}\quad \rho_\sigma=\half \dot{\sigma}^2 +V(\sigma),\hspace{.5em}\mbox{and}\quad V(\sigma)=\Lmd e^{-4\zeta\sigma},
 \label{lc_8} \\
\ddot{\sigma}+3H_*\dot{\sigma}+V'(\sigma)&=&0, \label{lc_9} \\
\dot{\rho}_* +4H_*\rho_*&=&0, \label{lc_10}
\end{eqnarray}
where a dot represents differentiation with respect to $t_*$, the cosmic time in the ECF to be discussed shortly, and we have
\begin{equation}
H_*=\frac{\dot{a}_*}{a_*}=\frac{da_* /dt_*}{a_*},\quad\mbox{with}\quad \dot{\sigma}=\frac{d\sigma}{dt_*}.
\label{lc_10a}
\end{equation}

The attractor solution is given by  
\begin{eqnarray}
a_*&=&t_*^{1/2}, \label{lc_11} \\
\sigma &=&\bar{\sigma}+\half \zeta^{-1}\ln t_*, 
\label{lc_12} \\
\rho_\sigma &=&\frac{3}{16}\zeta^{-2}t_*^{-2}, \label{lc_13} \\
\rho_*&=&\frac{3}{4}\left( 1-\frac{1}{4}\zeta^{-2} \right) t_*^{-2}, \label{lc_14}
\end{eqnarray}
with 
\begin{equation}
\Lmd e^{-4\zeta\bar{\sigma}} =\frac{1}{16}\zeta^{-2}.
\label{lc_14a}
\end{equation}

The physical condition $\rho_*>0$ applied to \reflef{lc_14}) is met under the same conditions \reflef{lc_15}), which correspond to the region as shown in Fig. 1 \cite{Bls}.

\mbox{}\\[.9em]
\hspace*{1.0em}
\bminip[h]{8.5cm}
\epsfxsize=7.0cm
\epsffile{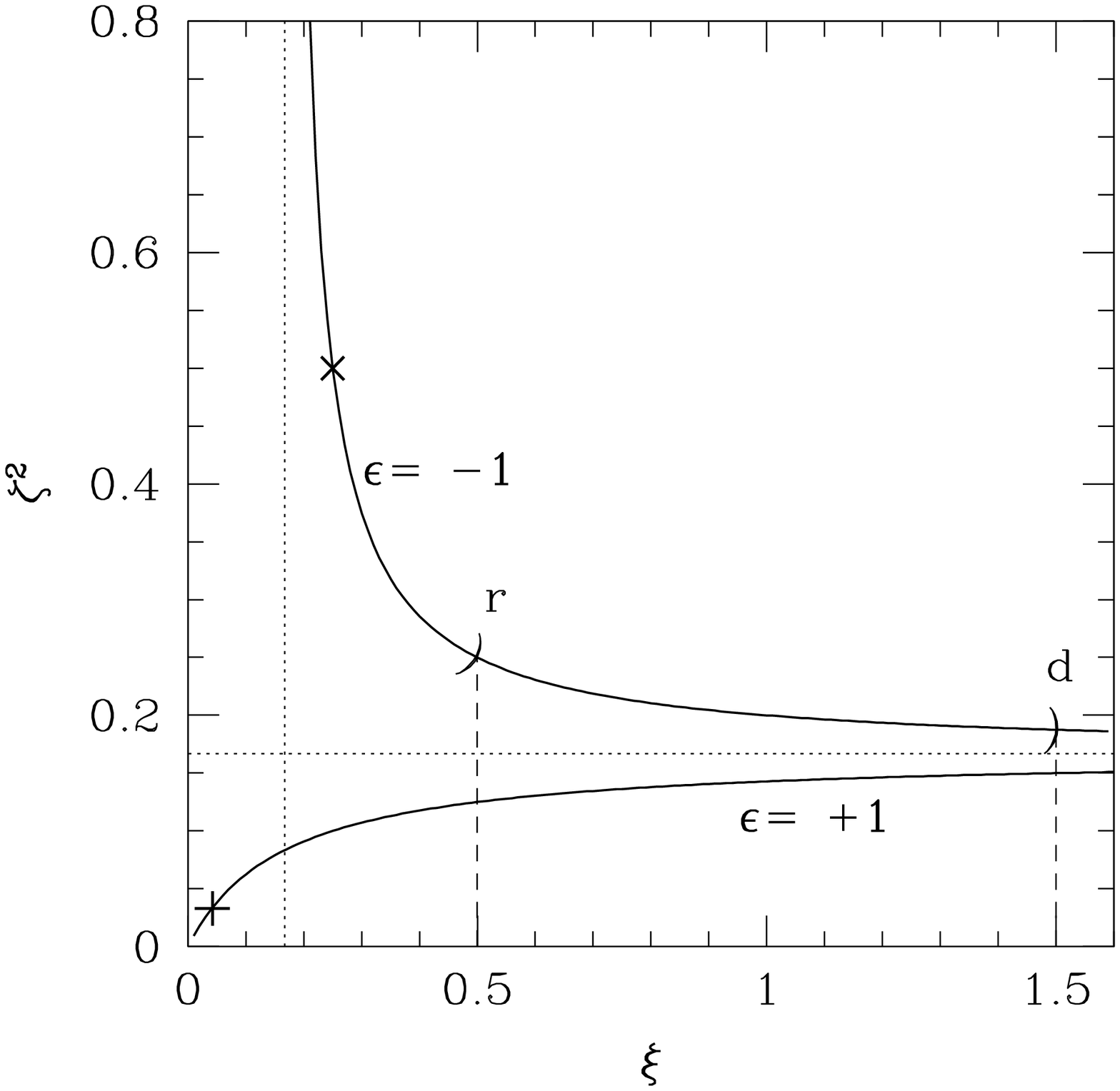}
\eminip
\hspace{.1em}
\bminip{8cm}
\mbox{}\\[-1.1em]
$\zeta^2$ given by \reflef{cftrI_6}) restricted to positive values is plotted as a function of $\xi >0$.  There are two branches for $\epsilon =\pm 1$ bounded by the two (dotted) lines $\xi =1/6$ and $\zeta^2 =1/6$, respectively.  The results from solar system experiments \cite{solar} correspond to the points, like the one marked by ``$+$" \cite{bd}, converging toward the origin $\xi =\zeta^2 =0$ with $\epsilon =+1$.  The symbol ``$)$" at the point $(\xi=0.5, \zeta^2=0.25)$ \reflef{lc_15}), marked by $r$, indicates that the portion of the curve with $\epsilon =-1$ to the upper-left is allowed for the radiation-dominated universe, while the point $(\xi =1.5,\zeta^2=3/16=0.1875)$ \reflef{lc_15D}) marked by $d$ is for the dust-dominated universe.  The symbol ``$\times$" indicates a prediction of string theory, $\epsilon =-1, \xi =1/4$ thus $\omega=-1$, and $\zeta^2 =1/2$ \reflef{gsw3}).

\eminip
\mbox{}\\[1.2em]

Now we relate the two solutions obtained for the JCF and the ECF.  First note that the  RHS of  \reflef{lc10_3a}) implies that $a$, which is {\em static} according to \reflef{lc_5}), represents an {\em expansion} if it is measured in units of a {\em shrinking} rod prepared by ${\cal M}^{-1}\sim t^{-1}$, as shown explicitly by the last equation of \reflef{lc_7b}).

Substituting \reflef{lc_5}) and \reflef{lc_7b}) into \reflef{lc10-1}), we find
\begin{equation}
a_* \sim t,
\label{lc_18b}
\end{equation}
which is compared with \reflef{lc_11}) to obtain
\begin{equation}
t\sim t_*^{1/2}.
\label{lc_18c}
\end{equation}
The technical details concerning the coefficients suppressed above for simplicity are given in Appendix A.

As stressed in \cite{dick2}, however, we recall that CFTR  is {\em different} from a general coordinate transformation which, unlike \reflef{lc10-2aa}), leaves the line element invariant.  For this reason, we continue to use the time coordinate $t$ unchanged  also in the ECF.  This is also connected with the relation
\begin{equation}
\hat{g}_{*00}=\Omega^2 g_{00}=-\Omega^2,
\label{lc_19}
\end{equation}
expected from applying \reflef{lc10-2}) to the $00$-component, together with $g_{00}=-1$.  Note that we have put a hat on the metric on the far-LHS in this intermediate stage, while, in the end, we need $g_{*00}=-1$ to define the cosmic time $t_*$, or the proper time in the coordinate system fixed to freely falling galaxies.  In order to clarify this puzzling situation, we apply a general coordinate transformation, $t \rightarrow t_*$, accompanied by the following transformation of the metric:
\begin{equation}
\hat{g}_{*00}=-\Omega^2 \rightarrow g_{*00}=\left( \frac{d t}{d t_*} \right)^2\hat{g}_{*00} = -1.
\label{lc_20b}
\end{equation}
It is this implication of a general coordinate transformation that we failed to make explicit in Ref.~\cite{cup} when we introduced differentiation with respect to $t_*$ tacitly in \reflef{lc_8}) through \reflef{lc_10a}).

From \reflef{lc_19}) and \reflef{lc_20b}) immediately follows 
\begin{equation}
\frac{dt}{dt_*}=\Omega^{-1}.
\label{lc_20bb}
\end{equation}
The consistency of this relation with \reflef{lc_18c}) is demonstrated in Appendix A.

The relation \reflef{lc_20bb}) could also have been obtained in parallel to deriving \reflef{lc10-1}), by applying \reflef{lc10-2}) to the $00$-component of the line element in the form 
\begin{equation}
-dt_*^2 = -\Omega^2 dt^2,
\label{lc_20bc}
\end{equation}
also giving another derivation of \reflef{lc_18c}) when combined with $\Omega \sim t$, as shown in \reflef{lc_7b}).  Further using this in \reflef{lc10-1}), together with \reflef{lc_5}), we arrive at \reflef{lc_11}), which is the same as the expansion law in the standard cosmology.  This supports the idea that the physical CF {\em happens} to be offered by the ECF.

We realize that the crucially important behavior $\phi \propto t$ in \reflef{lc_6}), and hence \reflef{lc_7b}), is a consequence of the simplest cosmological term on the RHS of \reflef{cftrI_1}), and hence of the constant  $\Lmd$ on the RHS of \reflef{dot_2}).  But this constancy might be relaxed by multiplying $\Lmd$  in \reflef{cftrI_1}) by a certain function $F(\varphi)$, as discussed in Appendix K of \cite{cup}, for example.  This will result in the corresponding change on the RHS of \reflef{lc_7b}) and on the RHS of \reflef{lc_11}) as well.   We could have been in a more advantageous position if the conclusion regarding the physical CF had been for a wider class of choices of the $\phi$-dependent cosmological term.  At this time, however, we will not discuss further details of the complications with full generality, except for the brief account given in Appendix B, in which the consistency of equations including \reflef{lc10-1}), \reflef{lc_20bb}) and \reflef{lc_11}) can be maintained only for the simplest choice of the $\phi$-independent cosmological term.  Also, in view of the fact that this choice entails a natural outcome in practical applications,  we are led to wonder whether we could employ an absolutely constant $\Lmd$ in some more fundamental law.


\subsection{Departure from the conformal dimensionless equation}

The question arises whether a CDlessE always follows given a quantity as long as it carries a dimension in the conventional sense, a length dimension, for example.  We find an exception immediately for $\sigma$ which has a length dimension of $-1$ but varies as $\ln t_*$ according to \reflef{lc_12}).  This makes it unlikely that we could find a JCF counterpart that  is related to it by certain power of $\Omega$.  However, we may consider the relation 
\begin{equation}
\Omega=e^{\zeta\sigma} =\xi^{1/2}\phi,
\label{lc_sgm_1a}
\end{equation}
which can be obtained by combining \reflef{cftrI_4}) and \reflef{lc_7b}).  Differentiating  this with respect to $t_*$, we obtain
\begin{equation}
\zeta\frac{d\sigma}{dt_*}e^{\zeta\sigma}=\xi^{1/2}\frac{dt}{dt_*}\frac{d\phi}{dt}.
\label{lc_sgm_1b}
\end{equation}
By using \reflef{lc_sgm_1a}) again together with \reflef{lc_7b}) and \reflef{lc_20bb}), we derive
\begin{equation}
\frac{d\sigma}{dt_*} = \zeta^{-1}\xi^{1/2}\Omega^{-2}\frac{d\phi}{dt}.
\label{lc_sgm_1}
\end{equation}
In fact the LHS is given by $(2\zeta)^{-1}t_*^{-1}$, according to \reflef{lc_12}), while we find $d\phi/dt= \kappa$ in view of \reflef{lc_6}).  With the help of \reflef{lc_18c}), we find that the LHS behaves as $\sim t^{-2}$, which is due to the factor $\Omega^{-2}$ on the RHS, thanks to \reflef{lc_7b}).  Thus, the validity of \reflef{lc_sgm_1}) is verified.  The exponent $-2$ agrees with the fact that the time derivative of the scalar field has a length dimension of $-2$.  Integrating this equation, we recover the correct relation \reflef{lc_sgm_1a}) between $\sigma$ and $\phi$, though somewhat indirectly.

Eliminating $\Omega$ by using \reflef{cftrI_7}), we may put \reflef{lc_sgm_1}) into the form of CDlessE,
\begin{equation}
{\cal M}_*^{-2}\dot{\sigma}={\cal M}^{-2}\dot{\phi},
\label{lc_sgm_2}
\end{equation}
suppressing the overall coefficient for the sake of simplicity (and we recall the relations $\dot{\sigma}=d\sigma/dt_*$ and $\dot{\phi}=d\phi/dt$).  The example here suggests the possible general feature that we may obtain the correct CDlessE by suitably avoiding the nonlinearity as exhibited by \reflef{lc_sgm_1a}), for example.

As another example of the failure of CDlessE, let us consider $H$ defined by the last equation in \reflef{dot_1}), together with the ECF counterpart, \reflef{lc_10a}).  By using \reflef{lc10-1}) and \reflef{lc_20bb}), we compute
\beqa
H_* &=&\frac{1}{a_*}\frac{dt}{dt_*}\frac{d}{dt}\left( \Omega a \right) \nnb\\
&=& \frac{1}{\Omega a}\Omega^{-1} \left( \Omega \dot{a}+\dot{\Omega}a \right) 
= \Omega^{-1}H+\frac{\dot{\Omega}}{\Omega^2},
\label{lc_27a}
\eeqa
where $\dot{\Omega}=d\Omega/dt$.  Suppose the second term on the far-RHS were absent, for the moment.  Then we would have the familiar relation
\begin{equation}
H_* =\Omega^{-1}H,
\label{lc_27b}
\end{equation}
indicating that $H$ would have a conformal dimension of $-1$, in agreement with the inverse of the length coming from differentiating with respect to $t$ and $t_*$ .  If this were true, we could combine \reflef{lc_27b}) with \reflef{cftrI_7}), arriving at
\begin{equation}
H_*{\cal M}_*^{-1}=H{\cal M}^{-1},
\label{lc_29a}
\end{equation}
precisely in the form of CDlessE.  However, this is erroneous, because the RHS vanishes, due to \reflef{lc_5}), whereas  the LHS equals $(1/2)t_*^{-1}$.  This disagreement can be traced back to the nonzero second term on the far-RHS of \reflef{lc_27a}), as we confirm
\begin{equation}
\frac{\dot{\Omega}}{\Omega^2}=\left( \xi^{1/2}\kappa \right)^{-1}t^{-2}=\frac{d\Omega /dt_*}{\Omega},
\label{lc_29b}
\end{equation}
where the result on the far-RHS has been obtained directly from the far-LHS, due to \reflef{lc_20bb}).

In this example, the failure of the simple CDlessE is due to the appearance of an inhomogeneous term on the far-RHS of \reflef{lc_27a}) depending on $\dot\Omega$.  This may suggest that naive assumption of the validity of CDlessE because of its origin in the intrinsic arbitrariness of units may be violated for a local $\Omega$, though \reflef{lc_sgm_2}) is an example of CDlessE, obviously a consequence of $\dot{\Omega}\neq 0$.  In both of the above two examples, the final results are obtained after investigating the field equations in a manner that goes beyond merely checking the conventional dimensions of the quantities in question.

Interestingly, a simple illustration of \reflef{lc_27a}) is found in the first equation of \reflef{lc_8}).  In fact, according to \reflef{lc_27a}) and $H=0$ from \reflef{lc_5}), $H_*^2$ on the LHS of \reflef{lc_8}) must behave as $(\dot{\Omega}/\Omega^2)^2 \sim (t^{-2})^2 =t^{-4}$, which is $t_*^{-2}$, due to \reflef{lc_18c}).  This is  precisely the form in which the RHS of the first equation of \reflef{lc_8}), $\rho_\sigma +\rho_*$, decays, according to \reflef{lc_13}) and \reflef{lc_14}).

As another aspect in which a complication may arise due to the nonlinear relation \reflef{lc_sgm_1a}), we discuss in Appendix C different descriptions we face in the two CFs concerning the possible nonzero mass of $\sigma$ and $\phi$.

\section{Cosmology in the dust-dominated universe}
\setcounter{equation}{0}

A new ingredient to be added to Section 2 is the mass terms of matter fields, conventionally given by the BD model, with the masses kept constant in the JCF.  Marked differences are found in Subsection 3.2 for the ECF, in which we find the disappointing result $a_*\sim t_*^{1/2}$, instead of the ordinary behavior, $a_*\sim t_*^{2/3}$.  Closely connected with this, we re-examine in Subsection 3.3 the behavior of  particle masses $m_*$ in the ECF, finding $m_*\sim t_*^{-1/2}$. We discuss the severity of the disagreement that this implies with the accepted idea of the expanding universe. We are then required to revise the conventional BD model, and in so doing (Subsection 3.4) we are fortunately rewarded by another simplification, global scale-invariance except for the $\Lmd$ term in the JCF Lagrangian.  This leads to the resurrection of the idea that ECF is qualified to be a physical CF, as in the radiation-dominated era.

\subsection{Jordan conformal frame}

In the preceding section, we discussed the radiation-dominated era, in which the masses of fundamental matter fields were ignored, assuming that the expansion of the universe had been ``measured" with respect to something other than the inverse of the particle masses, for example, with respect to the gravity length ${\cal M}^{-1}$, as defined in \reflef{cftrI_1a}) in the JCF, or to ${\cal M}_*^{-1} (=1)$ in the ECF.  Though this might be natural from a theoretical point of view, the universe in later epochs should be literally measured with respect to particle masses, mainly to the electron mass, which provides the unit of time for atomic clocks, as well as the atomic level differences to measure the redshifted wavelengths.  To facilitate this, we include particle masses.

We start with the BD model with \reflef{lc10_3b}).  Then, the  field equations in the JCF differ from those in the radiation-dominated universe by adding the term $\zeta^2 \rho$ to the RHS of \reflef{dot_2}) and replacing $4H\rho$ on the LHS of \reflef{dot_3}) with $3H\rho$, though the latter change has no effect for $H=0$, which is still a solution.  The corresponding changes in the solution are that $\kappa$ in \reflef{lc_6}) is replaced by $\tilde{\kappa}=\sqrt{(2\Lmd)/(4\xi +\epsilon)}$, and $\rho$ in \reflef{lc_7}) is replaced by $\tilde{\rho}=-2\Lmd (2\xi +\epsilon )/(4\xi +\epsilon )$.

We may also seek the solution with $a =t^{2/3}$ rather than $H=0$.  We find $\epsilon =+1$ and $\xi=3/2$, which imply $\zeta^2=3/20$.  This solution differs from the corresponding radiation-dominated solution for $a=t^{1/2}$ (obtained in Subsection 2.1 with $\xi =\infty,\ \zeta^2=1/6$) by too much to provide a coherent description of the universe.

\subsection{Einstein conformal frame}

In the ECF, in contrast to the JCF, the RHS of \reflef{lc_9}) acquires a source term, which produces the balancing term on the RHS of \reflef{lc_10}), basically to maintain the Bianchi identity (as elaborated in Appendix J of \cite{cup}),  given by
\beqa
\ddot{\sigma} +3H_* \dot{\sigma} +V'&=&\zeta\rho_*.  \label{clB_6b}\\
\dot{\rho}_*+3H_* \rho_*&=& -\dot{\sigma}\zeta\rho_*. \label{clB_6bb}
\eeqa
According to the same result as \reflef{lc_12}), we derive $\zeta\dot{\sigma}=(2t_*)^{-1}$, which happens to be the same as $H_*$, because we have $a_*=t_*^{1/2}$.   Rather unexpectedly, this behavior, instead of $a_*=t_*^{2/3}$, remains the same as in the case of radiation dominance.  We  find that the RHS of \reflef{clB_6bb}) plays the {\em same} role as the pressure term in the radiation dominated universe, leaving \reflef{clB_6bb}) as a whole the same as \reflef{lc_10}).  This is the way we understand the reason that \reflef{lc_11}) continues to apply to the case of dust dominance.

\subsection{Matter fields}

With the above preparation, we now consider a free neutral massive scalar field $\Phi$ as a simplified representative of microscopic fundamental fields.  The relevant Lagrangian in the JCF is given by
\begin{equation}
{\cal L}_\Phi =-\sqrt{-g}\left( \half g^{\mu\nu}\partial_\mu \Phi \partial_\mu \Phi +\half m^2 \Phi^2   \right),
\label{clPh_1}
\end{equation}
where $m$ is an absolute constant in the BD model, in which no $\phi$ is {\em required} to enter the matter Lagrangian to preserve WEP.

Under the CFTR given in \reflef{lc10-2}), we substitute $g_{\mu\nu}=\Omega^{-2}g_{*\mu\nu} $  into \reflef{clPh_1}) obtaining
\begin{equation}
{\cal L}_\Phi =-\sqrt{-g_*}\left( \half \Omega^{-2}g_*^{\mu\nu}\partial_\mu \Phi \partial_\mu \Phi +\half \Omega^{-4} m^2 \Phi^2   \right).
\label{clPh_2}
\end{equation}
Strictly speaking, the metric $g_{*\mu\nu}$ could be $\hat{g}_{*\mu\nu}$, as analyzed in \reflef{lc_19})-\reflef{lc_20bb}).  However, \reflef{clPh_2}) itself is invariant under any general coordinate transformation, including that discussed  in Subsection 2.2 and Appendix C concerning the cosmic time in ECF.  With this reservation in mind, we proceed with $\partial_0$, which may be $\partial /\partial t_*$.

The first term on the RHS of \reflef{clPh_2}) allows us naturally to define a new field $\Phi_*$ in such a way that it behaves like a standard field on Minkowski spacetime tangent to a manifold now endowed with $g_{*\mu\nu}$.  More specifically, we demand that $\Phi_*$, as a free field for the moment, have a canonical kinetic term
\begin{equation}
{\cal L}_{\Phi\overline{\rm KE}}=-\sqrt{-g_*}\half g^{\mu\nu}_* \partial_\mu \Phi_* \partial_\nu \Phi_*,
\label{clPh_2a}
\end{equation}
suggesting that we define $\Phi_*$ by
\begin{equation}
\Phi_* =\Omega^{-1}\Phi.
\label{clPh_3}
\end{equation}
In fact, substituting this into the first term on the RHS of \reflef{clPh_2}) yields
\begin{equation}
{\cal L}_{\rm \Phi KE}=-\sqrt{-g_*}\half g^{\mu\nu}_* {\cal D}_\mu \Phi_* {\cal D}_\nu \Phi_*,
\label{clPh_5}
\end{equation}
where
\begin{equation}
{\cal D}_\mu \Phi_* \equiv \left(\partial_\mu  +f_\mu \right)\Phi_*=\Omega^{-1}\partial_\mu\Phi,\quad\mbox{with}\quad f_\mu =\partial_\mu \ln \Omega =\zeta \partial_\mu\sigma.
\label{clPh_4}
\end{equation}
Equation \reflef{clPh_5}) certainly includes \reflef{clPh_2a}) as a free-field component, thus justifying \reflef{clPh_3}).

Now we substitute \reflef{clPh_3}) into the second term on the RHS of \reflef{clPh_2}), obtaining
\begin{equation}
{\cal L}_{\rm \Phi mass}=-\sqrt{-g_*}\half m_*^2 \Phi_*^2,
\label{clPh_6}
\end{equation}
where
\begin{equation}
m_* =\Omega^{-1}m
\label{clPh_7}
\end{equation}
can be interpreted as the mass of $\Phi_*$ in the ECF, as long as the resulting $m_*$ varies sufficiently slowly compared with any of the microscopic time-scales of the local system in which $\Phi_*$ plays a major role.  In a sense, $m_*$ is an ``instantaneous" mass.

We emphasize that we assumed no invariance under CFTR in the procedure described above.  In particular, the relation \reflef{clPh_7}) can be derived for a wider class of fields, including massless as well as massive gauge fields, i.e. fermions of spin 1/2 and 3/2, as elaborated in Appendix D, though the relation \reflef{clPh_3}) will be different for different spins, in general.  With all the microscopic mass scales obeying the same law, we expect the same to be true for all particles, including macroscopic objects.

Also, remarkably, a major part of the transformation factor $\Omega$ has been {\em absorbed} into the definition of $\Phi_*$ and $m_*$, leaving only a trace of the effect through the derivative of $\sigma$, as shown in the second equation of \reflef{clPh_4}).  This feature appears to be similar to, but not exactly the same as, what is called ``running units," or an additional rescaling to be applied in the ECF, as discussed in \cite{falao3}.  Our prescription also leaves the coupling of the matter fields ``normal," namely without being multiplied by powers of $e^{\zeta\sigma}$, in the sense discussed in \cite{cho}, as long as we stay at the classical level.

It is also worth noting that we can combine \reflef{clPh_7}) with \reflef{lc10-1}) to derive the CDlessE
\begin{equation}
a_*m_*=am.
\label{ccf_1_a}
\end{equation}
Due to \reflef{lc_5}) and \reflef{lc10_3b}) for the JCF mass, the RHS is time independent, and hence it gives
\begin{equation}
m_*^{-1} \sim a_*.
\label{ccf_1_b}
\end{equation}
This implies that the microscopic size expands in precisely the same way as the universe, which is completely incompatible with  the present fundamental understanding of the expansion of the universe, according to which   not only the microscopic structure but also the earth, stars and galaxies themselves remain decoupled from the cosmological expansion of inter-galactic distances.  In other words, $a_*$ is constant if we observe the cosmological expansion by measuring the inter-galactic distances in reference to microscopic rods.  This fatal defect, together with the behavior $a_*\sim t_*^{1/2}$ even in the case of dust dominance, as mentioned in Subsection 3.2, is discouraging.

Also, by combining \reflef{ccf_1_b}) with \reflef{lc_11}), we find
\begin{equation} 
m_* \sim t_*^{-1/2},
\label{clPh_8}
\end{equation}
which is in contradiction with the own-unit-insensitivity principle which holds when using atomic clocks in astronomical observations.  We have yet another reason why $m_*$ should be constant in any astronomical observation, as discussed below.

Consider how the distances to far away astronomical objects are estimated by measuring the redshift $z$: We compare atomic spectra in light emitted by them with the corresponding spectra prepared on the ground.  The basic {\em assumption} is that the wavelengths or the frequencies of the atomic transitions are basically the same in the far away emitters and the ground-based receivers.  This implies that the observed difference can result only from the redshift, or the spacetime geometry.  Without this assumption, we have no way to separate the geometric effect regarded as reflecting the expansion of the universe.  This can be compared with what we encounter in the Pound-Rebka experiment, in which radiation frequencies in the free-falling coordinate systems are the same for both of the emitter and receiver.  We also note that assuming the same frequencies in the heavens and the earth is nothing but assuming the same particle masses, particularly the electron mass, ignoring possible changes of the fine-structure constant, for the moment.  This is also in conformity with the own-unit-insensitivity principle that must hold when we use atomic clocks.

\subsection{Departure from the Brans-Dicke model}

We defined $m_*$ as a {\em local} constant, but it appears that it changes by too great an amount {\em globally} as a function of the cosmic time.  In other words, the instantaneous mass appears to change too rapidly during a much longer period of time.  Obviously, the real origin of this issue comes from \reflef{clPh_7}), with the condition \reflef{lc10_3b}).  To avoid this problem, we may try to instead impose 
\begin{equation}
m_* ={\rm const},
\label{clPh_9}
\end{equation}
in the ECF.  (We used the symbols $m_\Phi$ in (4.128) of \cite{cup} and $m_\natural$ in \cite{Bls}.)  This reminds us of \reflef{cftrI_7}), with ${\cal M}_*={\rm const}$,  possessing the same structure as \reflef{clPh_7}).  Exploiting this parallelism between ${\cal M}$ and $m$, we use \reflef{cftrI_1a}), which suggests  $m\propto \phi$,  replacing the mass term on the RHS of \reflef{clPh_1}) by
\begin{equation}
{\cal L}_{\Phi{\rm mass}}'=-\sqrt{-g}\frac{1}{2}f_\Phi^2 \phi^2\Phi^2,
\label{clPh_10}
\end{equation}
where $f_\Phi$ is a {\em dimensionless} coupling constant.  The absence of dimensional coupling constants can be extended to the whole matter Lagrangian, including any type of microscopic fields, as discussed in Appendix D.  This feature applies also to the gravitational part, consisting of the first two terms on the RHS of \reflef{cftrI_1}), but not to the $\Lmd$ term.  In spite of this limitation, it is amusing that our requirement \reflef{clPh_9}) is endowed with such a simple property, which also allows us an interpretation in terms of a global scale invariance at the level of the JCF Lagrangian.

We may interpret \reflef{clPh_10}) in terms of a time-dependent effective mass $m(t)$,
\begin{equation}
m(t) =f_\Phi\phi(t) \sim \kappa' f_\Phi t,
\label{clPh_10a}
\end{equation}
basically according to \reflef{lc_7b}),  with the modification stated toward the end of Subsection 3.1.

Applying a CFTR together with \reflef{clPh_3}), we obtain
\begin{equation}
{\cal L}_{\Phi{\rm mass}}'=-\sqrt{-g_*}\frac{1}{2}f_\Phi^2 \Omega^{-2}\phi^2 \Phi_*^2,
\label{clPh_11}
\end{equation}
which, thanks again to \reflef{lc_7b}),  gives the same result as \reflef{clPh_6}), 
\begin{equation}
{\cal L}_{\Phi{\rm mass}}'=-\sqrt{-g_*}\half m_*^2 \Phi_*^2,
\label{clPh_12}
\end{equation}
but now with a constant mass, as desired:
\begin{equation}
m_*=\xi^{-1/2}f_\Phi.
\label{clPh_12a}
\end{equation}
Because masses of ordinary matter particles are of order $\lsim {\rm GeV}\sim 10^{-18}$ in reduced Planck units, we expect a correspondingly small size of the coupling constants, $f_\Phi \lsim 10^{-18}$.

We find that the ``interaction term" \reflef{clPh_10}) breaks the BD assumption, thus inviting WEP violation in general.  On the other hand, however, the constant $m_*$ implies a scalar field $\sigma$ that is completely decoupled from $\Phi_*$, up to  derivative couplings arising from \reflef{clPh_4}).  In the limit of vanishing energy-momentum transfer, $\sigma$ has no interaction which could be responsible for the effect of WEP violation to be observed through the interaction among matter fields $\Phi_*$.  This approximate decoupling amounts to dropping the RHSs of \reflef{clB_6b}) and \reflef{clB_6bb}).  Then $3H_*\rho_*$ on LHS of \reflef{clB_6bb}) results in $a_*=t_*^{2/3}$, as usual.  This is certainly an encouraging sign for the ECF to be a physical CF, though we should keep in mind that this decoupling holds only at the classical level.  The coupling and the resulting WEP violation re-emerge at the quantum level, particularly through quantum anomalies, though the effects will be rather small, as discussed in Chapter 6 of \cite{cup}.

With this reservation, the revised model, called the ``scale-invariant model," represents the desired result for the dust-dominated universe in the ECF, in which we have a constant particle masses, $m_*$.  The scale factor $a_*(t_*)$ increases as $t_*^{2/3}$ if it is measured in reference to a microscopic rod provided by a constant $m_*^{-1}$.  The solution corresponding to \reflef{lc_11}) through \reflef{lc_14a})  is now given by 
\beqa
a_*&=&t_*^{2/3}, \label{lc_11D} \\
\sigma &=&\bar{\sigma}+\half \zeta^{-1}\ln t_*, 
\label{lc_12D} \\
\rho_\sigma &=&\frac{1}{4}\zeta^{-2}t_*^{-2}, \label{lc_13D} \\
\rho_*&=&\frac{4}{3}\left( 1-\frac{3}{16}\zeta^{-2} \right) t_*^{-2}, \label{lc_14D}
\eeqa
with
\begin{equation}
\Lmd e^{-4\zeta\bar{\sigma}} =\frac{1}{8}\zeta^{-2},
\label{lc_14Da}
\end{equation}
By demanding $\rho_*>0$, we find
\begin{equation}
\epsilon =-1,\quad\mbox{and}\quad \frac{1}{6}<\xi <\frac{3}{2} ,\quad\mbox{or}\quad \zeta^2>\frac{3}{16}=0.1875,
\label{lc_15D}
\end{equation}
which covers \reflef{lc_15}) for the case of radiation dominance, as also shown in Fig. 1.

In view of the fact that $m(t)$ defined by \reflef{clPh_10a}) is smaller for earlier times, the ECF solution for the radiation-dominated universe derived for the BD model is expected to carry over to the revised model.  We then accept \reflef{lc_11})-\reflef{lc_14a}) and \reflef{lc_11D})-\reflef{lc_14Da}) for the radiation- and the dust-dominated universes, respectively, as the ECF solution for the scale-invariant model as a whole.

In the same context, we may also inherit \reflef{lc_5})-\reflef{lc_7}) for the JCF solution in the radiation-dominated universe, based on the revised model.  As a nontrivial task, we then have to derive the JCF solution corresponding to the ECF solution for the case of dust dominance,  to be sketched below.

As we show in detail in Appendix E, the relation \reflef{lc_18c}) remains unchanged, but \reflef{lc_5}) is replaced by \reflef{cftd_6}):
\begin{equation}
a=t^{1/3}.
\label{lc_61}
\end{equation}
This appears to be the same as the consequence of $p/\rho=1$ in the language of a perfect fluid.  This interpretation is only tentative, however, because a rigorous analysis should be complicated by the violation of the conservation law $\nabla_\mu T^{\mu\nu}=0$ for the matter energy-momentum tensor, because of the appearance of $\phi$ in the matter Lagrangian, as in \reflef{clPh_10}).  Other relations which remain the same as in the case of radiation dominance include the asymptotic behavior $\phi \sim t$ as in \reflef{lc_6}), \reflef{lc_20bb}) and most of the general equations in Section 1.

The absence and presence of the mass terms in the JCF and the ECF, respectively, may be reinterpreted in terms of a spontaneous breaking of scale invariance in 4 dimensions, arising from the absence of dimensional constants in the Lagrangian in the JCF, except for the $\Lmd$ term, as discussed in detail in Chapters 6.1 and 6.2 of \cite{cup}.  The invariance is further violated explicitly through the loop effects, as also shown in Chapter 6.3 of \cite{cup}, including detailed analyses of quantum anomalies.  As has also been shown, the loop effects are typically proportional to the square of the coupling constant, obviously depending on specific properties of the matter fields.  Hence, they provide another source of WEP violation.  This feature, combined with a much smaller quark mass than the nucleon mass, results in a prediction of a somewhat weak coupling of the non-Newtonian field to nucleons, barely below the existing upper bound on the WEP violation, as discussed in Chapter 6.4 of \cite{cup}.

Also, it is likely that the electron mass is not constant in the ECF at the quantum level.  According to the own-unit-insensitivity principle, we must re-adjust the physical CF, but probably it must remain close to the classical ECF.  Another kind of readjustment may also be called for if the ratio $m_{\rm e}/m_{\rm p}$ turns out to be time dependent, as suggested in \cite{massratio}, because the time units of different atomic clocks are, strictly speaking, given by different reduced masses of the electron.

\section{Choosing conformal frames}
\setcounter{equation}{0}

Following our previous argument, we examine the two CFs from the points of view of two criteria, the Lagrangescape and the own-unit-insensitivity principle,  and we reach the conclusion that the ECF and the JCF are best interpreted in terms of observational and theoretical roles, respectively.    In Subsection 4.1 we explore how we can go back to earlier epochs, based on arguments mainly employing the ECF, which is acceptable as a physical CF based on the own-unit-insensitivity principle.  We then show in Subsection 4.2 how suited the JCF Lagrangescape is for investigating possible connections to more fundamental theories, like string theory and the Kaluza-Klein approach. Subsection 4.3 is devoted to a renewed discussion of the geodesic equations and WEP violation.

\subsection{Einstein conformal frame}

The present epoch is a dust-dominated universe.  In this case, the ECF appears to be a convenient choice, because the particle masses, denoted collectively by $m_*$, are time independent, while the scale factor $a_*$ is expanding as  $t_*^{2/3}$, at least to an approximation in which quantum effects are ignored for the moment.  We should recall that this is due to our revision of  BD's original model, in which $m$ in the JCF was chosen to be constant, and hence $m_*$ was too strongly time dependent in our approach of the ST$\Lmd$ cosmology.  In our scale-invariant model, on the other hand, we have instead a varying mass $m(t)$ in the JCF.

We maintain the relation \reflef{ccf_1_a}) in terms of the time-dependent $m$ introduced in \reflef{clPh_10a}).   Then we obtain the ``same" result if we measure $a$, which behaves as $\sim t^{1/3}$, according to \reflef{lc_61}), in reference to the shrinking microscopic length scale $m^{-1}\sim t^{-1}$, finding $am \sim t^{4/3}$, which agrees with $t_*^{2/3}$, due to \reflef{lc_18c}).  We point out here that $am$ in our model never exhibits exponential growth, in spite of the inclusion of $\Lmd$.  In this sense, the choice of the CF does not matter.  Nevertheless, choosing the ECF has the advantage that using $m_*^{-1}$ as a reference rod is automatically implied.  In other words, the own-unit-insensitivity principle is realized for atomic clocks, and for this reason, the ECF can be called a ``physical CF" in our model.  This property itself has its own merit, somewhat like the privileged status of the inertial frame in Newtonian mechanics in spite of the fact that general relativity allows us to study the phenomena in any of the coordinate frames.

In the JCF, or any other CF, on the other hand, we are always supposed to apply a readjustment to the standard we are using.  This can be done in accordance with a CDlessE like \reflef{lc10_3a}).  However, this may not always be true, as we illustrated in our examples in Subsection 2.3, based ultimately on the field equations derived from the Lagrangian in each CF.

Beyond this practical consideration, we wonder how the own-unit-insensitivity principle applies to the JCF.  In fact, in the radiation-dominated era, the only quantities which are time-dependent in principle but happen to be constant in our solution are $a$ and $\rho$.  Also, the constancy of one of them implies the same of the other, according to \reflef{dot_3}).  We may even say that the JCF is a CF in which the own-unit-insensitivity principle is realized for $a$ or $\rho$, which is an inconvenient way to view physics. Moreover, $a$ ceases to be constant in the dust-dominated universe, as shown in \reflef{lc_61}).  As a reminder, we note that the own-unit-insensitivity principle would have been realized for the particle mass $m$ in the dust-dominated universe if the BD model were correct, and thus  the JCF would be regarded as a physical CF.  We emphasize again that the accelerating universe forced us to introduce the large, constant  $\Lambda$, which resulted in the static radiation-dominated universe in the JCF, making it imperative to seek another CF as the physical one.

Even in epochs of radiation dominance,  we may still consider particle masses, as in the case of the primordial nucleosynthesis of light elements.  We maintain  the success of the standard theory based only on the scale-invariant model.  The BD model featuring \reflef{clPh_8}), on the other hand, would have suffered from a decrease in the particle masses predicted in non-relativistic quantum mechanics by as much as $\approx 70\%$ during the period of 100-1000 seconds, as discussed in Chapter 4.4.2 of \cite{cup}, particularly around (4.121).

But it may not make sense to continue to use atomic clocks when we go further back.  We now consider the question of what clock we could use to describe the evolution of an even earlier universe.  A possible choice is one based on the gravitational constant or gravity mass.  Fortunately, such a clock happens to be a continuous extension of our present atomic clocks, because the gravitational constant is time independent in the ECF both in dust- and radiation-dominated universes.  We warn that this continuation is not expected in the pure BD model, in which $m$ and ${\cal M}$ are constant only in the JCF and the ECF, respectively.   To summarize, from a practical point of view, the own-unit-insensitivity principle appears to select the ECF probably throughout the entire lifetime of the universe.

As one remarkable advantage of this wide perspective of the universe, we emphasize that \reflef{lc_13}) allows us to implement the ``scenario of a decaying cosmological constant," $\Lmd_{\rm eff}(t_*) \sim t_*^{-2}$.  We point out that the present age of the universe, $\sim 1.37\times 10^{10} {\rm y}$, is re-expressed as $\sim 10^{60}$ in units of the reduced Planck time, and thus today's $\Lmd_{\rm eff}$ is understood immediately to be of order $10^{-120}$.  This solves the fine-tuning problem mentioned at the beginning of Section 1: The present value of $\Lmd_{\rm eff}$ is small only because our universe is old.  This may also identify $\sigma$ with what is called the ``dark energy."  As detailed in Chapter 5.4.2 of \cite{cup}, the result is basically unchanged if we scrutinize the model by introducing another scalar field, $\chi$.  In fact, this also provides a {\em partial} solution to the coincidence problem \cite{cup,dodel}:  The acceleration, or the mini-inflation, we are presently observing is one of the repeated occurrences, and for each of  these, we have $\Lmd_{\rm eff}\sim t^{-2}$.

\subsection{Jordan conformal frame}

We next ask what significance we can attach to the JCF.  We first point out that the theory of closed strings is formulated in terms of a set of low-mass fields, including the massless scalar field $\Psi$, called a dilaton,  as a companion of the metric field in 26 dimensions.  The relevant part of the Lagrangian is given by 
\begin{equation}
{\cal L}_{\rm string}= \sqrt{-\bar{g}}e^{-2\Psi}\left(
\half \bar{R} +2 g^{\bar{\mu}\bar{\nu}}\partial_{\bar{\mu}} \Psi
\partial_{\bar\nu} \Psi 
 -\frac{1}{12} H_{\bar{\mu}\bar{\nu}\bar{\lambda}}
H^{\bar{\mu}\bar{\nu}\bar{\lambda}}
\right),
\label{gsw1}
\end{equation}
where the overbars indicate higher dimensions, while the last term on the RHS is for the field strength of a second-rank antisymmetric tensor field $B_{\bar{\mu}\bar{\nu}}$.  The above equation \reflef{gsw1}) is Eq. (3.4.58) of \cite{str} re-expressed in our notation and sign convention.

Introducing $\phi$ through the relation $\phi =2e^{-\Psi}$, the first two terms on the RHS of \reflef{gsw1}) can be put into the form
\begin{equation}
{\cal L}_{\rm st1}= \sqrt{-\bar{g}}\left( \half \xi \phi^2 \bar{R} - \half \epsilon 
g^{\bar{\mu}\bar{\nu}}\partial_{\bar{\mu}}\phi \partial_{\bar{\nu}}\phi \right),\label{gsw2}
\end{equation}
precisely as in \reflef{cftrI_1}) for the scalar-tensor theory in the JCF, where we have
\begin{equation}
\epsilon =-1,\quad \zeta^2 =\half, \quad \mbox{and}\quad \xi^{-1}=4, \quad\mbox{or}\quad \omega=-1.
\label{gsw3}
\end{equation}

We note that \reflef{gsw1}) is invariant under the simultaneous global transformations
\beqa
g_{\bar{\mu}\bar{\nu}}&\rightarrow& g'_{\bar{\mu}\bar{\nu}}=\lmd^2 g_{\bar{\mu}\bar{\nu}},
\label{gsw4a} \\
\Psi &\rightarrow& \Psi' =\Psi -\left( 1-D/2 \right)\ln \lmd, \hspace{.3em}\mbox{or}\hspace{.3em}\phi \rightarrow \phi' = \lmd^{1-D/2}\phi, \hspace{.3em}\mbox{with}\hspace{.3em} D=26,
\label{gsw4b} \\
B_{\bar{\mu}\bar{\nu}}&\rightarrow& B'_{\bar{\mu}\bar{\nu}}=\lmd^2 B_{\bar{\mu}\bar{\nu}}, \label{gsw4c}
\eeqa
as a descendant of the conformal invariance in 2-dimensional spacetime in which propagating strings reside.  Emphasizing this symmetry in the simplest manner yields the expression with the nonminimal coupling term.  We expect that the forms of \reflef{gsw1}) and \reflef{gsw2}) basically survive the process of dimensional reduction down to 4 dimensions (though this has never been demonstrated explicitly), arriving eventually at the scalar-tensor theory in the JCF, with the parameters as shown in Fig. 1.  We also expect that it is in this type of Lagrangian  that a large $\Lmd$ is derived, as seen from \cite{dz}.

String theory as a possible origin of the scalar field in the scalar-tensor theory was also studied by Damour and Polyakov \cite{DP}, who assumed, however, $\epsilon =+1$, without considering the question of how the sign reversal might have taken place in the dimensional reduction $D\rightarrow 4$.

Another candidate theory comes from the Kaluza-Klein theory in $D$ dimensions \cite{kk}, with the starting Lagrangian
\begin{equation}
{\cal L}_{\rm KK}= \half{\cal C}\sqrt{-\bar{g}}\bar{R},
\label{kk_1}
\end{equation}
where ${\cal C}$ is a constant.  As elaborated in Appendix A of \cite{cup}, the process of compactifying $n=D-4$ dimensions may be prepared by block-diagonalizing the metric tensor into the 4-dimensional part $g_{\mu\nu}$ and the $n$-dimensional ``internal space," assumed to be given by
\begin{equation}
g_{\alpha\beta}=A^2(x)\tilde{g}_{\alpha\beta},
\label{kk_2}
\end{equation}
where $\tilde{g}_{\alpha\beta}$ is a purely geometric metric whose indices run from 1 to $n$.

By integrating out the internal-space coordinates, we finally obtain the 4-dimensional Lagrangian for $n\neq 1$,
\begin{equation}
{\cal L}_4 =\sqrt{-g}\left( \half \xi \phi^2 -\half\epsilon g^{\mu\nu}\partial_\mu\phi \partial_\nu\phi +\half \left( \frac{1}{4}\frac{n}{n-1}\phi^2\right)^{1-2/n} \tilde{R} \right),
\label{kk_3}
\end{equation}
where the scalar field $\phi$ is defined by
\begin{equation}
\phi =2{\cal C}^{1/2}\sqrt{\frac{n-1}{n}}A^{n/2}.
\label{kk_4}
\end{equation}

The first two terms on the RHS of \reflef{kk_3}) are consistent with the corresponding terms  of the scalar-tensor theory with
\begin{equation}
\xi =\frac{1}{4}\frac{n}{n-1},\quad\mbox{with}\quad \epsilon =-1, \quad\mbox{for}\quad n>1,
\label{kk_5}
\end{equation}
whereas the last term on the RHS of \reflef{kk_3}) gives a cosmological constant $\Lmd$ if $n=2$, expressed in terms of the geometrical curvature scalar $\tilde{R}$, though this may have a sign that is opposite to that we would expect in \reflef{cftrI_1}). If $n \geq 3$, $\Lmd$ is multiplied by certain positive power of $\phi$, which is not favored from the point of view expressed at the end of Subsection 2.2 in selecting the ECF to be physical.

Obviously, $n=1$ or $D=5$ requires a separate analysis.  We have no kinetic energy term like the second term on the RHS of \reflef{kk_3}), corresponding to the choice $\epsilon =0$ with arbitrary $\xi$ and $\zeta^{-2}=6$, as we encountered in Subsection 2.1 with $a=t^{1/2}$.

We summarize the above exercises  by stating that the JCF provides an excellent Lagrangescape through which we may have  precious access to such fundamental theories as string theory and Kaluza-Klein compactification being manifested with a bare cosmological constant or its immediate extension.  It also seems worth emphasizing that both examples point to $\epsilon =-1$, as well as WEP violation, because of the appearance of the scalar field commonly in front of the matter Lagrangian.  The conclusion $\epsilon =+1$, with $\xi^{-1}=8$, is derived, on the other hand, in another 5-dimensional theory from the assumed two-sheeted spacetime \cite{saito}, as discussed in Chapter 1.2.4 of \cite{cup}, though without a unique result for $\Lmd$.

We still have no unique and final theoretical model for the desired absolutely constant $\Lmd$, as stated above, despite the support given in \cite{dz}, and also in the Kaluza-Klein theory with $D=6$.   It also is not obvious whether generating masses of ordinary light particles,  like quarks, leptons and gauge fields, in 4 dimensions, as expected from our scale-invariant model, fits the scenario envisioned in string theory.  We still hope that the lessons we have learned in the study of CFTR will serve as natural guides for the quest of an ultimate theory at more fundamental levels.

\subsection{Geodesic equations and WEP violation}

As another issue concerning the choice of the CF, it has been argued
\cite{bd,dick2,falao3,cho} that WEP is violated in the ECF even if we
start with the BD model in the JCF.  It is true that the geodesic
equation in the ECF acquires a nonzero RHS, which is, however, given by the
scalar field alone, independent of any specific properties of individual
matter objects.  Thus, the universal free-fall (UFF) which is the experimental manifestation of WEP, tested basically by the E\"{o}tv\"{o}s-type experiments is unaffected.  (For modern versions, see \cite {fisch} and \cite{eotwash}, for example.)

We start with the geodesic equation
\begin{equation}
\frac{Du^\mu}{D\tau}\equiv \frac{du^\mu}{d\tau}+\Gamma^\mu_{\ \nu\lmd}u^\nu u^\lmd =0, \quad\mbox{with}\quad u^\mu =\frac{dx^\mu}{d\tau},
\label{geods_1}
\end{equation}
in the JCF in the purely BD model.  Now, we apply CFTR \reflef{lc10-2}), obtaining
\begin{equation}
\frac{Du_*^\mu}{D\tau_*}= \left( f_\nu u_*^\nu \right)u_*^\mu -g_*^{\mu\nu}f_\nu \left( u_{*\lmd}u_*^\lmd \right),
\label{geods_2}
\end{equation}
as derived in (3.74) of \cite{cup}, where
\begin{equation}
f_\nu =\frac{\partial_\nu \Omega}{\Omega}, \quad\mbox{where}\quad
u_*^\mu =\frac{dx^\mu}{d\tau_*},\quad\mbox{with}\quad \frac{d\tau_*}{d\tau}=\Omega.
\label{geods_4}
\end{equation}
Thus, we have
\begin{equation}
u_\mu u^\mu =u_{*\mu} u_*^\mu.
\label{geods_5}
\end{equation}

If we consider massive matter particles, we can normalize \reflef{geods_5}) to be equal to $-1$.  The RHS of \reflef{geods_2}) then depends explicitly  only on the scalar field, irrespective of the specific properties of the individual matter particles, including the masses, and thus UFF is left intact.  In other words, every object at the same location in the static environment falls the same distance  in the same duration of time.  WEP holds as long as it is a principle supported by a physical process of UFF.  It is worth pointing out that \reflef{geods_2}) is directly concerned with the motion, or the trajectory $u_*^\mu (\tau_*)$, which enables us to verify UFF without appealing to the type of the Newtonian law of motion with mass and force as ingredients.

We may still study the changing mass on the basis of the equation
\begin{equation}
f_\nu=-\frac{\partial_\nu m_*}{m_*},
\label{geods_6}
\end{equation}
derived from \reflef{clPh_7}), combined with the first of \reflef{geods_4}).  This shows that the changing mass is solely determined by the scalar field, and it is common to any freely falling objects.   Using \reflef{geods_6}) together with $u_{*\lmd}u_*^\lmd =-1$, as mentioned above, we can re-express \reflef{geods_2}) as
\begin{equation}
\frac{d}{d\tau_*}\left( m_* g_{*\mu\rho} u_*^\rho \right) -m_*\half \left( \partial_\mu g_{*\nu\lmd}  \right) u_*^\nu u_*^\lmd  + \partial_\mu m_* =0,
\label{uff_9}
\end{equation}
which agrees with Eq. (69) of \cite{bd} and Eq. (27) of \cite{dick2}, with the last term called a ``nongeodesic" force.  However, this exerts a force exactly proportional to $m_*$ of each object and thus does not lead to the violation of UFF, which is consistent with the implication of the equivalent relation \reflef{geods_2}).

Suppose, on the other hand, we consider massless particles, such as photons, which move along a null geodesic with \reflef{geods_5}) vanishing.  Then the second term on the RHS of \reflef{geods_2}) vanishes, leaving the first term,  proportional to $u_*^\mu$.  The resulting equation is still a geodesic in a broader sense \cite{Wald}, and it can be brought back to the simplest type of geodesic without the RHS explicitly in the following manner \cite{misao}.

First, define $X$ by
\begin{equation}
\frac{d\ln X}{d\tau_*}=  f_\nu u_*^\nu,
\label{geods_7}
\end{equation}
from which we further define the {\em affine} parameter $\tilde{\tau}_*$ according to
\begin{equation}
d\tilde{\tau}_* =X d\tau_*.
\end{equation}
\label{geods_8}
It is then straightforward to put \reflef{geods_2}) without the second term on the RHS into the form of the homogeneous equation
\begin{equation}
\frac{D\tilde{u}_*^\mu}{D\tilde{\tau}_*}=0,\quad\mbox{with}\quad \tilde{u}_*^\mu=\frac{dx^\mu}{d\tilde{\tau}_*},  
\label{geods_9}
\end{equation}
which is the same as in the JCF.

Generally speaking, a nonzero RHS of the geodesic equation is only a necessary but not a sufficient condition for the breakdown of WEP.  No WEP violation emerges from none of its origin.  But the situation is different in the scale-invariant model.

In the ECF in the classical limit, as shown in Subsection 3.4, matter particles are decoupled from the scalar field $\sigma$, implying no RHS of the geodesic equation in the ECF.  The coupling emerges, however, if quantum effects are introduced.  Corresponding to the discussion toward the end of Subsection 3.4, we now have a non-trivial RHS of the geodesic equation.  We easily find theoretical differences  between charged and neutral particles, and between leptons and quarks.  However, an observationally significant deviation from UFF may be found only for particles including hadrons, for which complications may be unavoidable due to the strong interaction.


\section{Summary and concluding remarks}

We have presented examples of both useful ``equivalence" and its violation in results obtained from the Jordan and Einstein conformal frames, which are crucially important in the scalar-tensor theory.  This was made possible by {\em defining} equivalence and inequivalence in terms of CDlessE, like \reflef{lc10_3a}) or \reflef{lc_sgm_2}).  Equivalence is decided basically by consulting the field equations.  It may not be sufficient to consider only dimensionless ratios without examining the basic implications derived from the Lagrangian.

The argument made here does not consider whether if the underlying theory is conformally invariant.  Instead it is based only on the condition that the ``changes" due to the scalar field take place much more slowly than the microscopic changes.  The conclusions have been derived mainly from the ST$\Lmd$ cosmology, a theoretical model with a ``large" value of the cosmological constant  in the JCF.  For this reason, the scope of the discussion has been somewhat limited.  In particular, application to static, non-cosmological problems, for example, may require a re-analysis in terms of a different approach.  We nevertheless expect that the approach presented in this article will be extended to wider areas to be included for specific purposes.

As one of the points made in the present study, we reiterate that each CF is characterized with respect to two criteria, the Largangescape and the  own-unit-insensitivity principle.  From these criteria, we find that the JCF is useful for understanding how the real world is related to unification physics with a large $\Lmd$, offering a point of view beyond a more phenomenological ``quintessence" approach \cite{quint}.  On the other hand, the ECF, as a ``physical" conformal frame, applies to practical observations.  It does not seem fair to reject the JCF only because it is not ``physical."  The important role played by the JCF as a theoretical CF, as emphasized above, should be appreciated properly.

It also appears that the quintessence approach is confined to studies employing the ECF, even taking it for granted.  We point out, however, that we are always capable of moving to other CFs once we have a scalar field.  This scalar field is then used to produce a nonminimal coupling term.  We must find a reason that the ECF is chosen in most of practical applications.

We also urge the reader to recall that the ECF is qualified to be physical under the two conditions: (i) the term with $\Lmd$ in the JCF Lagrangian is a constant, though it can be so only asymptotically; (ii) the BD model in the JCF is revised so that WEP violation is allowed,  though there seem to be reasons why the effects are probably weak, as outlined near the end of Subsection 3.4.  More details should be worked out beyond the purely classical level, first through quantum anomalies.

We also believe it worthwhile to restate how the JCF is defined.  With the BD model, it used to be defined in terms of the matter Lagrangian, in which all the matter fields are decoupled from the scalar field $\phi$.  Given its necessary revision, we define it in terms of a CF in which all the matter fields enter the Lagrangian with couplings to $\phi$ through dimensionless constants. Needless to say, the ECF is defined by the Einstein-Hilbert term, and this implies that there is no time dependence of $G$, though quantum corrections will perhaps yield weak time dependence, as discussed at the end of Chapter 6.6.2 in \cite{cup}.

Further in connection with quantum effects, we remind the reader of the conclusion $\epsilon =-1$, as derived from our analysis of the ST$\Lmd$ cosmology.  This conclusion is also supported by string theory and the Kaluza-Klein compactification.  The observed small deviation from one of Eddington's parameters, $\gamma$, from 1 to fit the solar system experiments, $1-\gamma =4\zeta^2 \lsim 10^{-5}$ \cite{solar,yfms}, can be reconciled with the condition $\zeta^2 >1/6$ for $\epsilon =-1$, as illustrated in Fig. 1, only if the scalar field has a finite force-range, much shorter than the solar radius, for example, as has been suggested for the non-Newtonian force \cite{yfnonN}.

The expected finite force-range might be in contradiction with a uniformly decreasing potential $V(\sigma)\sim e^{-4\zeta\sigma}$, as shown in \reflef{cftrI_5}).  We must understand both of the requirements i.e., locally massive and globally massless natures of $\sigma$.  We believe that this can be done by means of the nontrivial behavior of the proper self-energy part of the scalar field \cite{yfSE}.  This approach appears to be related to another possible origin of $\Lmd$, the vacuum energy, in the sense of quantum field theory.  This has been a subject of much interest (see \cite{weinb}, for example), though it was not discussed in the present paper.

\section*{Acknowledgements}
I thank Kei-ichi Maeda, John Miller and Minoru Omote for valuable suggestions and comments.




\mbox{}\\[2.2em]
\appendix

\noindent
{\Large\bf Appendix}
\renewcommand{\thesection}{\Alph{section}}
\renewcommand{\theequation}{\Alph{section}.\arabic{equation}}
\section{ Multiplicative coefficients in the scale factors}
\setcounter{equation}{0}

Some of the dimensional constants appearing in the field equations used in Subsections 2.1 and 2.2  are given explicitly as follows:
\beqa
a&=& a_0 = {\rm const},   \label{clC_1} \\
a_*&=& a_{*0}t_*^{1/2},   \label{clC_2}
\eeqa
in \reflef{lc_5}) and \reflef{lc_11}), respectively.

We then replace \reflef{lc_18b}) and \reflef{lc_18c}), respectively, by
\beqa
a_*&=&a_0 \xi^{1/2}\kappa t, \label{clC_3} \\
t&=& \frac{a_{*0}}{a_0\xi^{1/2}\kappa}t_*^{1/2}.  \label{clC_4}
\eeqa
We then derive
\begin{equation}
\frac{dt}{dt_*}=\half \frac{a_{*0}}{a_0\xi^{1/2}\kappa} t_*^{-1/2} =\half \left( \frac{a_{*0}}{a_0\xi^{1/2}\kappa} \right)^2 t^{-1},
\label{clC_5}
\end{equation}
which should be equal to $\Omega^{-1}$, in accordance with \reflef{lc_20bb}).  Combining this with \reflef{lc_7b}), we obtain a constraint on the constants,
\begin{equation}
\left( \frac{a_{*0}}{a_{0}}\right)^2 = 2\xi^{1/2}\kappa.
\label{clC_7}
\end{equation}


\section{$\phi$-dependent cosmological term?}
\setcounter{equation}{0}

We start with the Lagrangian \reflef{cftrI_1}) with $\Lmd$ replaced by
$F(\varphi)\Lmd$, where $\varphi =(1/2)\xi \phi^2$.  The ``Einstein equation" is then
\begin{equation}
2\varphi G_{\mu\nu}=T_{\mu\nu}-g_{\mu\nu}F\Lmd+T_{\mu\nu}^\phi -2\left(g_{\mu\nu}\BBbox -\nabla_\mu \nabla_\nu   \right) \varphi.
\label{philmd_3}
\end{equation}

Varying the basic Lagrangian with respect to $\phi$, we obtain
\begin{equation}
\xi\phi R+\epsilon \BBbox \phi -\xi\phi F'\Lmd =0,
\label{philmd_4}
\end{equation}
where
\begin{equation}
F' \equiv \frac{dF}{d\varphi} =\frac{dF}{d\phi}\frac{d\phi}{d\varphi}=\xi^{-1}\phi^{-1}\frac{dF}{d\phi}.
\label{philmd_5}
\end{equation}
Multiplying \reflef{philmd_4}) by $\phi$ yields
\begin{equation}
2\varphi R+\epsilon \phi \BBbox \phi=2\varphi F'\Lmd.
\label{philmd_6}
\end{equation}

We next multiply \reflef{philmd_3}) by $g^{\mu\nu}$ to obtain
\begin{equation}
-2\varphi R +\epsilon \left( \partial\phi \right)^2 = T - 4F\Lmd -6\BBbox \varphi.
\label{philmd_7}
\end{equation}
Eliminating $2\varphi R$ from \reflef{philmd_6}) and \reflef{philmd_7}), we find
\begin{equation}
\BBbox \varphi =\zeta^2 T -4\zeta^2 \left( F-\half \varphi F' \right)\Lmd.
\label{philmd_9}
\end{equation}

As in Subsection 2.1, we set $T=H=0$ in the radiation-dominated universe, obtaining
\begin{equation}
\ddot{\varphi} =4\zeta^2 \left( F-\half \varphi F' \right)\Lmd.
\label{philmd_9a}
\end{equation}

We next check the consistency of the equations
\begin{equation}
a_* = \Omega a =\Omega,
\label{philmd_19}
\end{equation}
for \reflef{lc10-1}),
\begin{equation}
dt_*=\Omega dt,\quad\mbox{namely}\quad \frac{dt}{dt_*}=\Omega^{-1},
\label{philmd_20}
\end{equation}
for \reflef{lc_20bc}) and 
\begin{equation}
a_* =t_*^{1/2},
\label{philmd_20a}
\end{equation}
for \reflef{lc_11}) which was obtained for the ECF solution.  We also add
\begin{equation}
\Omega =\xi^{1/2}\phi,
\label{philmd_20b}
\end{equation}
implied by \reflef{cftrI_1a}) and \reflef{lc10-2a}).

We attempt to find solutions for the simplified assumption
\begin{equation}
F(\varphi)=\varphi^\alpha,
\label{philmd_10}
\end{equation}
by which \reflef{philmd_9a}) can be put into the form
\begin{equation}
\ddot{\varphi}=4\zeta^2 \left( 1-\frac{\alpha}{2} \right)\varphi^\alpha \Lmd.
\label{philmd_13}
\end{equation}

Obviously, the RHS vanishes for $\alpha=2$, resulting in
\begin{equation}
\varphi = {\rm const}\times t, \quad\mbox{then}\quad \phi =t^{1/2},
\label{philmd_13a}
\end{equation}
where in the second equation, we have suppressed the multiplicative coefficient which is nonessential.  This implies, through \reflef{philmd_20b}) and \reflef{philmd_19}), the following:
\begin{equation}
a_*=\Omega =t^{1/2}.
\label{philmd_23}
\end{equation}
Comparing this with \reflef{philmd_20a}), we should have
\begin{equation}
t=t_*,
\label{philmd_24}
\end{equation}
leading to
\begin{equation}
\frac{dt}{dt_*}=1,
\label{philmd_25}
\end{equation}
which cannot be identified with \reflef{philmd_20}).  In this way, we find an internal inconsistency for $\alpha=2$ among the equations that comprise the mathematical ingredients of the present analysis of CFTR.

Equation \reflef{philmd_13}) shows that there is another exceptional choice, $\alpha=1$, giving the solution
\begin{equation}
\phi =\xi^{-1/2}\Omega =\exp \left( \zeta\sqrt{\Lmd /2}\:t  \right).
\label{philmd_26}
\end{equation}
Substituting this into \reflef{philmd_20b}) and equating the result with \reflef{philmd_19}) with the help of \reflef{philmd_20a}), we find
\begin{equation}
t=\ln \left( t_*^{1/2} \right)=\half \ln t_*,
\label{philmd_27}
\end{equation}
from which we derive
\begin{equation}
\frac{dt}{dt_*}=t_*^{-1}=\Omega^{-2},
\label{philmd_28}
\end{equation}
where we have used \reflef{philmd_20a}) again and \reflef{philmd_19}) on the far-RHS.  We again find disagreement with \reflef{philmd_20}).

Finally, we consider the case in which $\alpha$ is equal to neither of 1 nor 2.  We assume
\begin{equation}
\varphi =t^\beta, \quad\mbox{or}\quad \phi=t^{\beta/2}.
\label{philmd_14}
\end{equation}
Substituting this into \reflef{philmd_13}) yields
\begin{equation}
\ddot{\varphi}=\beta (\beta-1)t^{\beta-2}=4\zeta^2 \left( 1-\frac{\alpha}{2} \right) t^{\alpha\beta}\Lmd,
\label{philmd_15}
\end{equation}
from which follow
\begin{equation}
\beta =\frac{2}{1-\alpha},\quad\mbox{and}\quad \frac{1+\alpha}{(1-\alpha)^2(2-\alpha)}=\zeta^2\Lmd.
\label{philmd_16}
\end{equation}
Using \reflef{philmd_14}) in \reflef{philmd_20b}), and equating the result again with  \reflef{philmd_20a}) through \reflef{philmd_19}), we obtain
\begin{equation}
t^{\beta}=t_*.
\label{philmd_30}
\end{equation}
We then find
\begin{equation}
\frac{dt_*}{dt} =\beta t^{\beta -1},
\label{philmd_31}
\end{equation}
which must be the same as $\Omega$, which equals the second equation of \reflef{philmd_14}).  This requires
\begin{equation}
\beta-1=\frac{\beta}{2},\quad\mbox{or}\quad \beta =2.
\label{philmd_33}
\end{equation}
According to the first equation of \reflef{philmd_16}), this implies $\alpha=0$, bringing us back to a constant cosmological term.

\section{Mass term of $\phi$ or $\sigma$}
\setcounter{equation}{0}

Here, we study how to modify the formulation in order to give a nonzero mass to the scalar field.  We start with \reflef{BD_1}), but suppress the term with $\Lmd$ for simplicity in this appendix, though we may replace $T_{\mu\nu}$ by $T_{\mu\nu} -\Lmd g_{\mu\nu}$ whenever necessary.  Thus, we consider
\begin{equation}
\BBbox \varphi =\zeta^2 T,
\label{appb_1}
\end{equation}
with
\begin{equation}
\varphi =\half \xi\phi^2 = \half \Omega^2=\half e^{2\zeta\sigma},
\label{appb_2}
\end{equation}
where use has been made of \reflef{lc_sgm_1a}).   Below, we conformally transform \reflef{appb_1}).

We begin with the LHS, using \reflef{appb_2}):
\beqa
\BBbox\varphi &=&\frac{1}{\sqrt{-g}}\partial_\mu \left( \sqrt{-g}g^{\mu\nu}\partial_\nu \varphi \right) =
\Omega^4\frac{1}{\sqrt{-g_*}}\partial_\mu \left( \sqrt{-g_*}g_*^{\mu\nu}
\Omega^{-2}\left(  \Omega^2 \zeta \partial_\nu \sigma \right)\right) \nnb\\
&=&\zeta \Omega^4\frac{1}{\sqrt{-g_*}}\partial_\mu  \left(\sqrt{-g_*}g_*^{\mu\nu}\partial_\nu \sigma \right)
=\zeta \Omega^4 \BBbox_*\sigma.
\label{appb_5}
\eeqa
In order to analyze the RHS of \reflef{appb_1}), we start with the standard argument   concerning the matter energy-momentum tensor in each CF.  Then, according to (3.52) of \cite{cup}, we derive 
\begin{equation}
T_{\mu\nu}=\Omega^2 \hat{T}_{*\mu\nu},
\label{appb_21}
\end{equation}
where the hat on $T_{\mu\nu}$ indicates that we are still in an intermediate stage in which the conformally transformed metric is given by 
\begin{equation}
\hat{g}_{*00}=-\Omega^2,\quad\mbox{with}\quad \Omega =\frac{dt_*}{dt},
\label{appb_21_1}
\end{equation}
as shown by \reflef{lc_19}) and \reflef{lc_20bb}), where $t_*$ represents the cosmic time in the ECF.  We then proposed to apply a general coordinate transformation to realize the relation 
\begin{equation}
g_{*00}=\Omega^{-2}\hat{g}_{*00} =-1,\quad\mbox{together with}\quad T_{*00}=\Omega^{-2}\hat{T}_{*00},
\label{appb_21_2}
\end{equation}
so that
\begin{equation}
\hat{g}_*^{00}\hat{T}_{*00}=g_*^{00}T_{*00}.
\label{appb_21_4}
\end{equation}
This is readily extended to
\begin{equation}
\hat{T}_* \equiv \hat{g}_*^{\mu\nu}\hat{T}_{*\mu\nu} =g_*^{\mu\nu}T_{*\mu\nu}\equiv T_*,
\label{appb_21_5}
\end{equation}
as long as we are dealing with a general coordinate transformation, including the special choice discussed above of non-vanishing components only for $\mu=\nu=0$, but strictly speaking, we have the added assumption $g_{i0}=0$ allowed in the FRW geometry.  Note that requiring $t_*$ to be a cosmic time has no effect on the {\em trace} of $T_{*\mu\nu}$.

Combining \reflef{appb_21_5}) with \reflef{appb_21}) and \reflef{lc10-2}), with $g_{\mu\nu}$ replaced by $\hat{g}_{\mu\nu}$, we thus obtain
\begin{equation}
T\equiv g^{\mu\nu}T_{\mu\nu}=\Omega^2\hat{g}_*^{\mu\nu}\Omega^2 \hat{T}_{*\mu\nu}=\Omega^4 \hat{T}_*=\Omega^4 T_*.
\label{appb_25}
\end{equation}
Substituting this and \reflef{appb_5}) into \reflef{appb_1}), we arrive at
\begin{equation}
\BBbox_*\sigma =\zeta T_*,
\label{appb_6}
\end{equation}
in agreement with Eq. (3.54) of \cite{cup}.

Now, let us modify this to
\begin{equation}
\left( \BBbox_* -\mu^2 \right)\sigma =\zeta T_*,
\label{appb_7}
\end{equation}
where $\mu$ is assumed to have come from the loop diagrams of other matter fields.  Note that $\sigma$ couples to matter fields in the ECF not only in the BD model but also in the scale-invariant model at the quantum level, as discussed in Subsection 3.4.  The first term on the LHS is readily brought back to $\zeta^{-1}\Omega^{-4}\BBbox\varphi$ again by using \reflef{appb_5}).  We then consider the second term on the LHS of \reflef{appb_7}) by inverting \reflef{appb_2}):
\begin{equation}
-\mu^2\sigma =-\mu^2\zeta^{-1}\half \ln (2\varphi).
\label{appb_8}
\end{equation}
Multiplying \reflef{appb_7}) by $\zeta\Omega^4$, we obtain
\begin{equation}
\BBbox\varphi -2\mu^2 \varphi^2 \ln (2\varphi)=\zeta^2 T,
\label{appb_9}
\end{equation}
where we have used \reflef{appb_25}) and $\Omega^4=4\varphi^2$, obtained from \reflef{appb_2}).  The second term on the LHS is quite different from the ordinary mass term $-\mu^2\varphi$.

There may be a simplification in the weak-field limit, however, with which we have
\begin{equation}
\phi \approx \xi^{-1/2}\left( 1+\zeta\tilde{\sigma} \right),\quad\mbox{hence}\quad \varphi \approx \half +\zeta\tilde{\sigma},
\label{appb_10}
\end{equation}
where we have identified the vacuum expectation value of $\varphi$ to be 1/2, corresponding to the Einstein-Hilbert term.  Substituting \reflef{appb_10}) into the second term on the LHS of \reflef{appb_9}), and keeping only the linear term, we find that \reflef{appb_9}) takes the form
\begin{equation}
\left( \BBbox -\mu^2  \right) \tilde{\sigma}=\zeta T,
\label{appb_11}
\end{equation}
which is what we would find by replacing $\sigma$ in \reflef{appb_7}) with $\tilde{\sigma}$ and replacing $\BBbox_*$ and $T_*$ with $\BBbox$ and $T$.


In place of the above argument, we may also attempt, alternatively, to start by modifying \reflef{appb_1}) to
\begin{equation}
\left( \BBbox-\mu^2 \right)\varphi =\zeta^2 T.
\label{appb_12}
\end{equation}
Then, using \reflef{appb_5}) again in the first term on the LHS, we find $\zeta\Omega^4 \BBbox_*\sigma$.  In the second term on the LHS multiplied by $\zeta^{-1}\Omega^{-4}$, we substitute the last term of \reflef{appb_2}), thus obtaining
\begin{equation}
-\mu^2 \zeta^{-1}\Omega^{-4}\half \Omega^2=-\mu^2 \zeta^{-1}\half e^{-2\zeta\sigma}.
\label{appb_13}
\end{equation}
We hence obtain
\begin{equation}
\BBbox_* \sigma -\zeta^{-1}\half\mu^2 e^{-2\zeta\sigma}=\zeta T_*.
\label{appb_14}
\end{equation}
To linear order in $\sigma$, we thus find
\begin{equation}
\left( \BBbox_* +\mu^2 \right)\sigma = \half \zeta^{-1}\mu^2 +\zeta T_*.
\label{appb_15}
\end{equation}

The inhomogeneous term $(2\zeta)^{-1}\mu^2$ on the RHS can be removed by shifting $\sigma$ as
\begin{equation}
\sigma \rightarrow \sigma +\half \zeta^{-1},
\label{appb_16}
\end{equation}
which is, however, effective only at linear order.  The wrong sign of the term containing $\mu^2$ on the LHS of \reflef{appb_15}) can be changed only by changing the sign in \reflef{appb_12}).  However, these procedures are unnatural, probably because, unlike $\sigma$ in the ECF, $\phi$ or $\varphi$ in the JCF is not a diagonalized field, making the assumption \reflef{appb_12}) somewhat invalid.   In this respect, the assumption \reflef{appb_7}) appears better, though loops of the same type as those considered  for $\mu^2$ might contribute another term, giving a new mixing coupling between $\sigma$ and the spinless portion of the metric field.  At this time, we only expect that this introduces another diagonalization process with which the above simple-minded result remains self-consistent.


\section{Masses of other types of matter fields}
\setcounter{equation}{0}

In this appendix, we show that \reflef{clPh_7}) derived for the spinless field $\Phi$ can also be obtained in a wider class of fields.

Consider a free real tensor field $A_{\rho_1\rho_2 \cdots \rho_n}$ of rank $n$ in $D$ dimensions with either symmetrized or antisymmetrized indices.  The case $n=0$ is that of a spinless field, while $n=1$ corresponds to a gauge vector field.  In more detailed analysis of higher-spin fields, we have to decompose the field into irreducible representations of the Lorentz group before identifying the result with a physically acceptable value of the spin.  At this time, however, we avoid those details, focusing on simple relations for the fields of given $n$ and $D$.

The JCF Lagrangian for the kinetic energy term is given, apart from a multiplicative numerical factor depending on $n$, by
\begin{equation}
{\cal L}_{ {\rm KE}}=-\sqrt{-g}\:g^{\mu\nu} g^{\rho_1\sigma_1}\cdots g^{\rho_n\sigma_n}\left(\nabla_{\mu}A_{\rho_1 \cdots \rho_n}\right)\left( \nabla_{\nu}A_{\sigma_1 \cdots \sigma_n}\right),
\label{cftra_5_1}
\end{equation}
where $\nabla_\mu$ is the covariant derivative.  As in \reflef{clPh_2a}), we focus upon the derivative terms of highest order, 
\begin{equation}
{\cal L}_{\overline{\rm KE}}=-\sqrt{-g}\:g^{\mu\nu} g^{\rho_1\sigma_1}\cdots g^{\rho_n\sigma_n}\left(\partial_{\mu}A_{\rho_1 \cdots \rho_n}\right)\left( \partial_{\nu}A_{\sigma_1 \cdots \sigma_n}\right).
\label{cftra_5_1a}
\end{equation}
Next, we substitute
\begin{equation}
g_{\mu\nu}=\Omega^{-2}g_{*\mu\nu},\quad g^{\mu\nu}=\Omega^{2}g_*^{\mu\nu},\quad \sqrt{-g}=\Omega^{-D}\sqrt{-g_*}
\label{cftra_5_2}
\end{equation}
into \reflef{cftra_5_1a}), obtaining
\begin{equation}
{\cal L}_{\overline{\rm KE}}=-\sqrt{-g_*}\:\Omega^{-D}g_*^{\mu\nu} \Omega^{2(n+1)}g_*^{\rho_1\sigma_1}\cdots g_*^{\rho_{n}\sigma_{n}}\left( \partial_{\mu}A_{\rho_1 \cdots \rho_{n}}\right)\left( \partial_{\nu}A_{\sigma_1 \cdots \sigma_{n}}\right).
\label{cftra_5_3}
\end{equation}

With
\begin{equation}
d=\frac{D}{2},
\label{cftra_5_3a}
\end{equation}
for an even value of $D$, there is a special case in which
\begin{equation}
d=n+1,
\label{cftra_5_4}
\end{equation}
for which $\Omega$ disappears in \reflef{cftra_5_3}).  We then find 
\begin{equation}
{\cal L}_{\overline{\rm KE}}=-\sqrt{-g_*}\:g_*^{\mu\nu} g_*^{\rho_1\sigma_1}\cdots g_*^{\rho_{n}\sigma_{n}}\left( \partial_{\mu}A_{*\rho_1 \cdots \rho_{n}}\right)\left(\partial_{\nu}A_{*\sigma_1 \cdots \sigma_{n}}\right),
\label{cftra_5_5}
\end{equation}
where
\begin{equation}
A_{*\rho_1\rho_2 \cdots \rho_n}=A_{\rho_1\rho_2 \cdots \rho_n},
\label{cftra_5_6}
\end{equation}
which shows the conformal invariance for ${\cal L}_{\overline{\rm KE}}$.  In 2 and 4 dimensions, this is the case for the spinless and the unit spin fields, respectively.  We develop the formulation without restriction to such massless conformally invariant fields.

If \reflef{cftra_5_4}) does not hold, we introduce the quantity $A_{*\rho_1\rho_2 \cdots \rho_n}$ as
\begin{equation}
A_{*\rho_1 \cdots \rho_n}=\Omega^{n+1-d}A_{\rho_1 \cdots \rho_n},
\label{cftra_5_7}
\end{equation}
which yields
\begin{equation}
\Omega^{n+1-d}\partial_\mu A_{\rho_1 \cdots \rho_n}=\left[ \rule[-.05em]{0em}{.9em}\partial_\mu -(n+1-d)f_\mu \right] A_{*\rho_1 \cdots \rho_n} \equiv {\cal D}_\mu A_{*\rho_1 \cdots \rho_n},
\label{cftra_5_8}
\end{equation}
where
\begin{equation}
f_\mu= \partial_\mu \ln \Omega =\zeta \partial_\mu \sigma,
\label{cftra_5_9}
\end{equation}
precisely as in \reflef{clPh_4}).

Further covariantizing ${\cal D}_\mu$ by ${\cal D}_{*\mu}$ we put \reflef{cftra_5_1}) into the form
\begin{equation}
{\cal L}_{\rm KE}=-\sqrt{-g_*}\:g_*^{\mu\nu} g_*^{\rho_1\sigma_1}\cdots g_*^{\rho_{n}\sigma_{n}} \left( {\cal D}_{*\mu}A_{*\rho_1 \cdots \rho_{n-1}}\right)\left({\cal D}_{*\nu}A_{*\sigma_1 \cdots \sigma_{n}}\right),
\label{cftra_5_11}
\end{equation}
which is a natural extension of \reflef{cftra_5_5}).   We add that we still have the freedom to apply a general coordinate transformation, like that mentioned in Subsection 2.2 and in Appendix C.

The mass term is given typically by
\begin{equation}
{\cal L}_{\rm mass}=-\sqrt{-g}\:m^2  g^{\rho_1\sigma_1}\cdots g^{\rho_{n}\sigma_{n}}A_{\rho_1 \cdots \rho_{n}}A_{\sigma_1 \cdots \sigma_{n}},
\label{cftra_5_12}
\end{equation}
where $m$ is a constant in the JCF according to the BD model.  By substituting from \reflef{cftra_5_2}) and \reflef{cftra_5_7}), we derive
\begin{equation}
{\cal L}_{\rm mass}=-\sqrt{-g_*}\:m_*^2  g_*^{\rho_1\sigma_1}\cdots g_*^{\rho_{n}\sigma_{n}}A_{*\rho_1 \cdots \rho_{n}}A_{_*\sigma_1 \cdots \sigma_{n}},
\label{cftra_5_13}
\end{equation}
where we define the instantaneous mass $m_*$ as
\begin{equation}
m_*=\Omega^{-1}m.
\label{cftra_5_14}
\end{equation}

Note that $\Omega$ always appears in \reflef{cftra_5_14}) with the exponent $-1$ for any $n$, in agreement with the conventional value, $-1$, the inverse length dimension.  On the other hand, \reflef{cftra_5_7}) shows that the field has the conformal dimension $n+1-d$, which agrees with the conventional value, $1-d$, for flat spacetime, only for $n=0$.  Obviously, \reflef{cftra_5_14}) introduces an explicit breaking of the conformal invariance whenever we have a massive field.

The ultimate origin of the behavior described by  \reflef{cftra_5_14}) can be traced back to the difference between \reflef{cftra_5_1}) and \reflef{cftra_5_12}): $g^{\mu\nu}$ appears in the former but not in the latter.  We also reiterate that we {\em defined} $m_*$ as a factor appearing in \reflef{cftra_5_13}), together with defining $A_{*\rho_1\cdots \rho_n}$ by \reflef{cftra_5_7}), quite independently of the question of whether or not the conformal invariance is valid.  On the other hand, the relation \reflef{cftra_5_14}) is, by itself, not useful in practical applications, but it is useful if we add the relation $\Omega \sim t$, which appears to be most natural from an overall theoretical point of view, as stated above.

We also add that the Higgs' mechanism provides the vector field mass basically through the relation $m =gv$, where $g$ and $v$ are the gauge coupling constant and the vacuum expectation value of a scalar field.  We combine this with the familiar process through which $v$ results from a negative mass-squared and a quartic coupling constant, as discussed in Chapter 6.2 of \cite{cup}, in which $v$ obeys the rule \reflef{clPh_7}).  In this way, we finally find that the gauge field mass $m$ behaves according to the same law.  At the more technical level, the Proca Lagrangian derived for a massive unit-spin field automatically yields a condition of the form $g^{\mu\nu}\nabla_\mu A_\nu =0$, which leads to the sum of \reflef{cftra_5_1}) and \reflef{cftra_5_12}).

Extending the analysis to a spin $1/2$ fermion is straightforward.  Consider the kinetic energy term    for a Dirac field, as given by
\begin{equation}
{\cal L}_{\rm KE}= -b\psibar\left( b^{\mu i}\gamma_i D_\mu  \right)\psi,
\label{cftra_5_21}
\end{equation}
where the Greek and Latin indices are for the curved and flat spacetimes, respectively.  We confine our consideration to 4 dimensions for simplicity in what follows.  The Latin indices are raised or lowered by means of the flat metric $\eta_{ij}, \eta^{ij}$.  We also have
\begin{equation}
g_{\mu\nu}=b_{\mu i}b_{\nu}^i,\quad g^{\mu\nu}=b^\mu _ib^{\nu i},\quad \sqrt{-g}=b,
\label{cftra_5_22}
\end{equation}
in which the ``tetrad" $b_{\mu i}$ is the ``square-root" of the metric $g_{\mu\nu}$, with the corresponding relations for $b^\mu_ i$ \cite{utiyama}.  Also, $\gamma_i$ is the usual constant (spcetime independent) Dirac matrix.

The covariant derivative is defined by
\begin{equation}
D_\mu =\partial_\mu +\frac{1}{4}\omega^{ij}_{\:\;,\mu}\half\left[ \gamma_i, \gamma_j \right].
\label{cftra_5_23}
\end{equation}
The spin connection $\omega^{ij}_{\:\;,\mu}$ can be ignored in the current simplified analysis, in which we focus upon the terms of the highest order derivative of $\psi$.  By comparing \reflef{lc10-2}) and \reflef{cftra_5_22}), we find
\begin{equation}
b_{\mu i}=\Omega^{-1}b_{*\mu i}, \quad b^\mu _i=\Omega b^\mu_{* i}, \quad b=\Omega^{-4}b_*.
\label{cftra_5_24}
\end{equation}
Further, dropping the terms of the spin connections for simplicity, we then obtain
\begin{equation}
{\cal L}_{\overline{\rm KE}}= -b_* \Omega^{-3}{\psibar}\left( b_*^{\mu i} \Omega \gamma_i \partial_\mu  \right)\psi.
\label{cftra_5_25}
\end{equation}

The quantity $\Omega^{-3}$ is absorbed into $\psi_*$ and $\psibar_*$, and we thus obtain
\begin{equation}
{\cal L}_{\overline{\rm KE}}= -b_* \psibar_*\left( b_*^{\mu i} \gamma_i \partial_\mu  \right)\psi_*,
\label{cftra_5_26}
\end{equation}
where
\begin{equation}
\psi_* =\Omega^{-3/2}\psi,\quad \psibar_* =\Omega^{-3/2}\psibar.
\label{cftra_5_27}
\end{equation}
Note that \reflef{cftra_5_26}) appears invariant only up to terms of $\partial_\mu\sigma$.  We find, in fact, that the terms with spin connections fully included are completely conformally  invariant  under the torsion-free condition, as shown explicitly in Appendix F of \cite{cup}.

The mass term 
\begin{equation}
{\cal L}_{\rm mass}=-bm\psibar \psi
\label{cftra_5_28}
\end{equation}
is then re-expressed as
\begin{equation}
{\cal L}_{\rm mass}=-b_*m_*\psibar_* \psi_*,
\label{cftra_5_29}
\end{equation}
with
\begin{equation}
m_* =\Omega^{-1}m,
\label{cftra_5_30}
\end{equation}
which is the same as \reflef{clPh_7}) for the bosonic fields.

Now, let us turn to the spin-$3/2$ Rarita-Schwinger field with the vector-spinor field $\psi_\rho$ and the Lagrangian \cite{dzrs}
\begin{equation}
{\cal L}_{\rm RS}=-b\psibar_\rho \left(\gamma^{\rho\mu\sigma}D_\mu +mg^{\rho\sigma} \right)\psi_\sigma,
\label{cftra_5_31}
\end{equation}
where $\gamma^\mu =b^\mu_i \gamma^i$, with $\gamma^{\rho\mu\sigma}$ the antisymmetrized product of $\gamma^\rho \gamma^\mu \gamma^\sigma$, while $D_\mu$ is the  covariant derivative defined by \reflef{cftra_5_23}).  By re-expressing the $\gamma$'s with Greek indices in terms of those with Latin indices, we find
\begin{equation}
b\gamma^{\rho\mu\sigma}=i\epsilon^{\rho\mu\sigma\nu}\gamma_5\gamma_\nu,
\label{cftra_5_31a}
\end{equation}
where $\gamma_5 \equiv i\gamma_0\gamma_1 \gamma_2 \gamma_3=-\rho_1$ with the Latin $\gamma$'s, is a spacetime independent numerical matrix, while $\epsilon^{\rho\mu\sigma\nu}$ is the Levi-Civita tensor, taking the values $0$ or $\pm 1$, {\em without} a factor of the determinant $b$.  We then put the first term on the RHS of \reflef{cftra_5_31}) into the form
\begin{equation}
{\cal L}_{\rm KE}=-i\half \epsilon^{\rho\mu\sigma\nu}  \psibar_\rho\gamma_5\gamma_\nu D_\mu \psi_\sigma.
\label{cftra_5_31b}
\end{equation}
Focusing upon the derivative of $\psi_\sigma$, we simply replace $D_\mu$ by $\partial_\mu$.  When we move to the ECF, only $b_{\nu i}$ in $\gamma_\nu$ is multiplied by $\Omega^{-1}$, which is absorbed into the vector-spinor fields:  
\begin{equation}
\psi_{*\sigma} =\Omega^{-1/2}\psi_\sigma,\quad \psibar_{*\sigma} =\Omega^{-1/2}\psibar_\rho.
\label{cftra_5_31c}
\end{equation}

The mass term is given by
\begin{equation}
{\cal L}_{\rm mass}=-bm g^{\rho\sigma}\psibar_\rho \psi_\sigma= -b_*\Omega^{-1}m g_*^{\rho\sigma}\psibar_{*\rho} \psi_{*\sigma},
\label{cftra_5_34}
\end{equation}
from which we immediately reach the relation \reflef{cftra_5_30}).




\section{Jordan frame solution in the dust-dominated universe}
\setcounter{equation}{0}

In the radiation-dominated universe in the JCF, we ignored the masses of matter fields, which we include by adding the term $-\zeta^2 T \sim \zeta^2 \rho$ to the RHS of \reflef{dot_2}).  Below, we show that a consistent solution can be obtained for $\rho ={\rm const}$, which exhibits the same behavior as that of the existing term, $4 \Lmd$.  In this context we start with \reflef{lc_6}) and \reflef{lc_7b}),
\begin{equation}
\phi =\xi^{-1/2}\Omega =\hat{\kappa} t,
\label{cftd_21}
\end{equation}
but with a different coefficient $\hat{\kappa}$.

Assume
\begin{equation}
a=t^\gamma.
\label{cftd_1}
\end{equation}
Substituting this, together with \reflef{lc_11D}), into \reflef{lc10-1}) yields
\begin{equation}
\Omega =t_*^{2/3}t^{-\gamma}.
\label{cftd_2}
\end{equation}
We combine this with another general equation, \reflef{lc_20bb}), obtaining
\begin{equation}
\frac{dt_*}{dt}=t_*^{2/3}t^{-\gamma},
\label{cftd_3}
\end{equation}
from which follows
\begin{equation}
t_*=t^{3(1-\gamma)}.
\label{cftd_4}
\end{equation}
Using this in \reflef{cftd_2}), we obtain
\begin{equation}
\Omega =t^{2-3\gamma}.
\label{cftd_5}
\end{equation}
Comparing this with \reflef{cftd_21}) gives
\begin{equation}
2-3\gamma =1,\quad\mbox{hence}\quad \gamma=\frac{1}{3}, \quad\mbox{or}\quad H=\frac{1}{3}t^{-1}.
\label{cftd_6}
\end{equation}
With this result, \reflef{cftd_4}) now reduces to
\begin{equation}
t_*=t^2,
\label{cftd_6a}
\end{equation}
which is the same as \reflef{lc_18c}) for the case of radiation dominance.

We must investigate how the appearance of $\phi^2$ on the RHS of \reflef{clPh_10}) affects the RHS of \reflef{dot_3}), with $4H\rho$ on the LHS replaced by $3H\rho$.  According to (D.15) of Ref.~\cite{cup}, with $\phi$ there replaced by $\phi^2$, we find
\begin{equation}
\nabla_\mu T^{\mu\nu}=\phi^2\nabla^\nu \left( \half f_\Phi^2 \Phi^2\right),
\label{cftd_7}
\end{equation}
which can be put into the form
\begin{equation}
\dot{\rho}+3H\rho =-\phi^2\partial_t\left( \half f_\Phi^2 \Phi^2\right).
\label{cftd_8}
\end{equation}

We simplify the analysis here by assuming
\begin{equation}
\rho =m^2 \Phi^2,
\label{cftd_9}
\end{equation}
with
\begin{equation}
m=f_\Phi \phi =f_\Phi\hat{\kappa} t, 
\label{cftd_10}
\end{equation}
following the revision proposed in \reflef{clPh_10}).  The assumed constancy of \reflef{cftd_9}) then requires
\begin{equation}
\Phi =\Phi_0 t^{-1},
\label{cftd_10a}
\end{equation}
with a constant $\Phi_0$.   Hence, we have
\begin{equation}
\rho=\left( f_\Phi \hat{\kappa} \Phi_0 \right)^2.
\label{cftd_10b}
\end{equation}
From this, together with \reflef{cftd_6}) and \reflef{cftd_10a}), we find that \reflef{cftd_8}) is satisfied.

From \reflef{cftd_21}) we obtain
\begin{equation}
\varphi =\half \xi \hat{\kappa}^2 t^2.
\label{cftd_10c}
\end{equation}
Substituting this, as well as \reflef{cftd_6}), into \reflef{dot_1}), we finally obtain
\begin{equation}
\frac{\Lmd}{\hat{\kappa}^2} =\frac{7}{3}\xi -f_\Phi^2 \Phi_0^2 -\half \epsilon.
\label{cftd_23}
\end{equation}


\end{document}